%% file: manuscript_diborides.tex
\documentclass[reprint, superscriptaddress, amsmath,amssymb, aps]{revtex4-2}
\usepackage{xcolor}
\usepackage{soul}
\newcommand{\add}[1]{{\textcolor{black}{#1}}}
\usepackage{graphicx}
\usepackage{xcolor}
\usepackage{dcolumn}
\usepackage{bm}
\usepackage{hyperref}
\usepackage{gensymb} 
\hypersetup{
    colorlinks=true,
    linkcolor=green,
    urlcolor=cyan,
}
\usepackage{todonotes}

\begin{document}

\preprint{njp Computational Materials}

\title{On energetics of allotrope transformations in transition-metal diborides via plane-by-plane shearing}

\author{Thomas Leiner}
\email{thomas.leiner@unileoben.ac.at}
\affiliation{Department of Materials Science, Montanuniversität Leoben, Leoben, Austria}
\author{Nikola Koutná}
\affiliation{Institute of Materials Science and Technology, TU Wien, Vienna, Austria}
\affiliation{Department of Physics, Chemistry, and Biology (IFM), Link\"{o}ping University, Link\"{o}ping, Sweden}
\author{\add{Jozef Janovec}}
\affiliation{\add{Centro de Física de Materiales-MPC CSIC-UPV/EHU, San Sebastián, Spain}}
\affiliation{\add{Institute of Materials Science and Engineering, Faculty of Mechanical Engineering, Brno University of Technology, Brno, Czech Republic}}
\author{\add{Martin Zelený}}
\affiliation{\add{Institute of Materials Science and Engineering, Faculty of Mechanical Engineering, Brno University of Technology, Brno, Czech Republic}}
\author{Paul H. Mayrhofer}
\affiliation{Institute of Materials Science and Technology, TU Wien, Vienna, Austria}
\author{David Holec}
\affiliation{Department of Materials Science, Montanuniversität Leoben, Leoben, Austria}

\date{\today}

\begin{abstract}
Transition metal diborides crystallize in the $\alpha$, $\gamma$, or $\omega$ type structure, in which pure transition metal layers alternate with pure boron layers stacked along the hexagonal [0001] axis.
Here we view the prototypes as different stackings of the transition metal planes and suppose they can transform from one into another by a displacive transformation.
Employing first-principles calculations, we simulate sliding of individual planes in the group IV--VII transition metal diborides along a transformation pathway connecting the $\alpha$, $\gamma$, and $\omega$ structure.
Chemistry-related trends are predicted in terms of energetic and structural changes along a transformation pathway, together with \add{the} mechanical and dynamical stability of the different stackings.
Our results suggest that MnB$_2$ and MoB$_2$ possess the overall lowest sliding barriers among the investigated TMB$_2$s.
Furthermore, we discuss trends in strength and ductility indicators, including Young's modulus or Cauchy pressure, derived from elastic constants.
\end{abstract}


\maketitle

\section{Introduction}
Transition metal diborides (TMB$_2$s) are a vibrant research topic in application-oriented coating developments \cite{fuger2019influence,fuger2022anisotropic,palisaitis2022nature,palisaitis2021unpaired,vsroba2020structure,paul2021plastic} and represent promising materials for usage in extreme environments, including ultra-high temperatures and severe mechanical loads.
Among their attractive properties are chemical stability and inertness, high melting point, high hardness, good electrical and thermal conductivity, corrosion, and erosion resistance as well as high wear and thermal-shock resistance \cite{magnuson2021review,mitterer1997borides}. 
TMB$_2$s with extraordinary characteristics include TaB$_2$ and ReB$_2$---showing hardness values up to 46--49\;GPa \cite{liu2022superhard} and 48\;GPa \cite{ultra_high_hardness_ReB2}, respectively---or e.g. MgB$_2$, which is a superconductor with a critical temperature of 39\;K \cite{nagamatsu2001superconductivity}.
Among the most widely researched diborides is TiB$_2$\cite{hellgren2022synthesis,bakhit2020improving}, valued for its high chemical stability and high melting point ({$3500$\;K}), high hardness ({$24$\;GPa}) and chemical inertness, as well as electric resistivity and thermal conductivity \cite{munro2000material}. 
High hardness ({$26$\;GPa}) and low electrical resistivity is reported also for WB\textsubscript{2}, applicable as a conductor under extreme conditions \cite{properties_WB2}.

TMB$_2$s are known to commonly crystallize in layered structures with hexagonal symmetry, most often in the AlB$_2$-type phase ($\alpha$, space group $\#$191--P6/mmm) which is typical for diborides of early transition metals \cite{magnuson2021review}. 
Furthermore, the ReB$_2$-type phase ($\gamma$, space group $\#194$--P6$_3$/mmc) has been reported for ReB$_2$~\cite{La_Placa1962-pj} and the  WB$_2$-type phase ($\omega$, space group $\#$194--P6$_3$/mmc) can be stabilized for late TMB$_2$s \cite{moraes2018ab}.
All three phases---$\alpha$, $\gamma$, and $\omega$---can be viewed as layered structures that alternate hexagonal nets of pure TM atoms and layers of pure B atoms (arranged in honeycombs), stacked along the $c$-axis.
Using the standard labeling for stacking of hexagonal planes, the arrangement of the metal atoms can be described as A-A-A-A, A-B-A-B, and A-A-B-B  stacking sequence for the $\alpha$, $\gamma$, and $\omega$ phase, respectively. 
The boron sheets between the metal planes come in two configurations: either flat (H) as in the $\alpha$ phase, or puckered (K) as in the $\gamma$ phase. 
Using this nomenclature, first introduced by \citet{Kiessling1947-sb}, the 3 structural prototypes are described as:
\begin{tabbing}
    \quad \= $\alpha$:\quad \= \dots-A-H-$\overbrace{\mbox{A-H-A-H-A-H-A-H}}$-A-H-\dots\ ,\\
    \> $\gamma$: \> \dots-B-K-$\overbrace{\mbox{A-K-B-K-A-K-B-K}}$-A-K\dots\ , \\
    \> $\omega$: \> \dots-B-K-$\overbrace{\mbox{A-H-A-K-B-H-B-K}}$-A-H\dots\ 
     .
\end{tabbing}
The structures, therefore, do not only differ by the stacking of the metal planes (A, B) but also by the geometry of the boron planes.
While these are all flat in the $\alpha$ structure, they are all puckered in the $\gamma$ structure. 
The $\omega$ phase contains alternating flat and puckered boron sheets.
It is interesting to observe that the flat H configuration of B planes always appears when the surrounding metal planes have the same stacking (i.e. A-A or B-B). 
We note that the above formalism allows also for other stackings, e.g., \dots-A-H-A-K-B-H-B-K-C-H-C-K-\dots (structure of Mo$_2$B$_5$~\cite{La_Placa1962-pj}) which are, however, not the focus of the present work.

The fact that the three prototypes differ primarily by the stacking of the TM planes provokes the idea that they can transform from one into another by shearing/sliding individual planes, i.e. by a displacive transformation. 
This corresponds to the $(0001)[1\bar100]$ slip, previously identified as the origin of easy plasticity of ReB$_2$~\cite{Zhang2010-ah} and later found active also in ZrB$_2$~\cite{Hunter2016-ee}.
We note that other slip systems, such as $(1\bar{1}00)[11\bar{2}3]$ or  $(0001)[11\bar{2}0]$, may be operative for different TMB$_2$s depending on temperature~\cite{Paul2021-pd}.

In this work\add{,} we simulate sliding of individual planes in TMB$_2$s using first-principles calculations. 
We aim \add{to provide} chemistry-related trends for 12 group IV--VII transition metal diborides in terms of their stability, as well as energetic and structural changes along a transformation pathway connecting all three prototypes, namely, AAAA $\to$ \add{BAAA} $\to$ ABAB $\to$ ABBA $\to$ AAAA.
Here AAAA, ABAB, and ABBA correspond to the metal plane stackings in the $\alpha$, $\gamma$, and $\omega$ structures, respectively.

\section{Methods}
The calculations were performed using the Vienna \textit{ab-initio} Simulation Package (VASP) \cite{Kresse1996Efficient} using the projector-augmented wave \add{(PAW)} method and a plane wave basis set \cite{Kresse1999From}. 
The exchange-correlation effects were treated with the aid of generalized gradient approximation (GGA) functionals from Perdew-Burke-Ernzerhof (PBE) \cite{perdew1996generalized}. The $\Gamma$-centered $k$-point mesh was automatically generated with a length parameter of 50\;\AA \, while the plane waves cut-off energy was set to 500\;eV. 

\add{The simulation cells consist of four metal layers and four boron layers and, therefore, 12 atoms each.}
Geometric constraints were set using the GADGET code by  \citet{buvcko2005geometry}.
Specifically, all structures were allowed to relax in the $a$- and $c$-direction, while keeping the lattice angles $\alpha$, $\beta$ and $\gamma$ at 90\degree\ and 120\degree, respectively, to preserve the hexagonal crystal symmetry. 
Boron atoms were allowed to move freely, whereas metal atoms could move only along the [0001] ($c$) direction.
All systems (including CrB$_2$ and MnB$_2$) were treated as non-magnetic.
The energy barrier between stackings $\mathcal{S}_1$ (e.g. AAAA) and $\mathcal{S}_2$ (e.g. BAAA) a is calculated as
\begin{equation}
    E({\cal S}_1\to{\cal S}_2)=
    \begin{cases}
    E_{\text{tot}}^{\text{max}}-E_{\text{tot}}^{\text{min}},& \text{if } E_{\text{tot}}({\cal S}_1)<E_{\text{tot}}({\cal S}_2)\\
    E_{\text{tot}}^{\text{max}}-E_{\text{tot}}({\cal S}_1),  & \text{otherwise},
    \end{cases}
\label{EQ: E-barrier}
\end{equation}
where $ E_{\text{tot}}^{\text{max}}$ ($ E_{\text{tot}}^{\text{min}}$) is the maximal (minimal) total energy along the ${\cal S}_1\to{\cal S}_2$ pathway, and $E_{\text{tot}}({\cal S}_1)$, $E_{\text{tot}}({\cal S}_2)$ are total energies of $\mathcal{S}_1$ and $\mathcal{S}_2$, respectively.
We note that Eq.~\eqref{EQ: E-barrier} allows to resolve directionally-dependent barriers, i.e. $E({\cal S}_1\to{\cal S}_2)$ is generally different from $E({\cal S}_2\to{\cal S}_1)$ .

To assess mechanical stability and predict trends in elastic properties, the stress-strain method \cite{le2002symmetry,le2001symmetry,Yu2010-vr} was used to calculate fourth-order elasticity tensors, mapped onto symmetric $6\times6$ matrices of elastic constants, $\{C_{ij}\}$, via Voigt's notation.
Subsequently, positive definiteness of the $\{C_{ij}\}$ matrix (equivalent to the positivity of its minimal eigenvalue) served as a necessary and sufficient criterion for the mechanical stability of the corresponding structure \cite{mouhat2014necessary}. 
Imposing the macroscopic symmetry, elastic matrices were projected on those of a hexagonal system thus yielding five independent elastic constants ($C_{11}$, $C_{12}$, $C_{13}$, $C_{33}$, and $C_{44}$).
The polycrystalline Young's modulus, $E=9BG/(3B+G)$ was calculated using the Hill's average of the bulk, $B$, and shear modulus, $G$ \cite{nye1985physical,hill1952elastic}.

Furthermore, dynamical stability was addressed through calculating phonon spectra, using the Phonopy package \cite{togo2008first} with a 4$\times$4$\times$1 (192-atom) replica of the fully relaxed diboride simulation cell\add{, using the small displacements method with the default displacement of 0.01 \AA.}
\add{To analyze the chemical bonding, we take advantage of the crystal orbital Hamilton population (COHP)
\cite{dronskowski1993crystal}---a tool that weights the density of states (DOS) by the elements of the Hamilton matrix---calculated within the LOBSTER package \cite{maintz2013analytic} capable of extracting chemical information from plane-wave wave function by its transformation onto a local basis set \cite{maintz2016lobster}.}

\section{Results and discussion}
\subsection{Transformation energy landscape}
\begin{figure*}[h!t!]
	\centering
    \includegraphics[width=2\columnwidth]{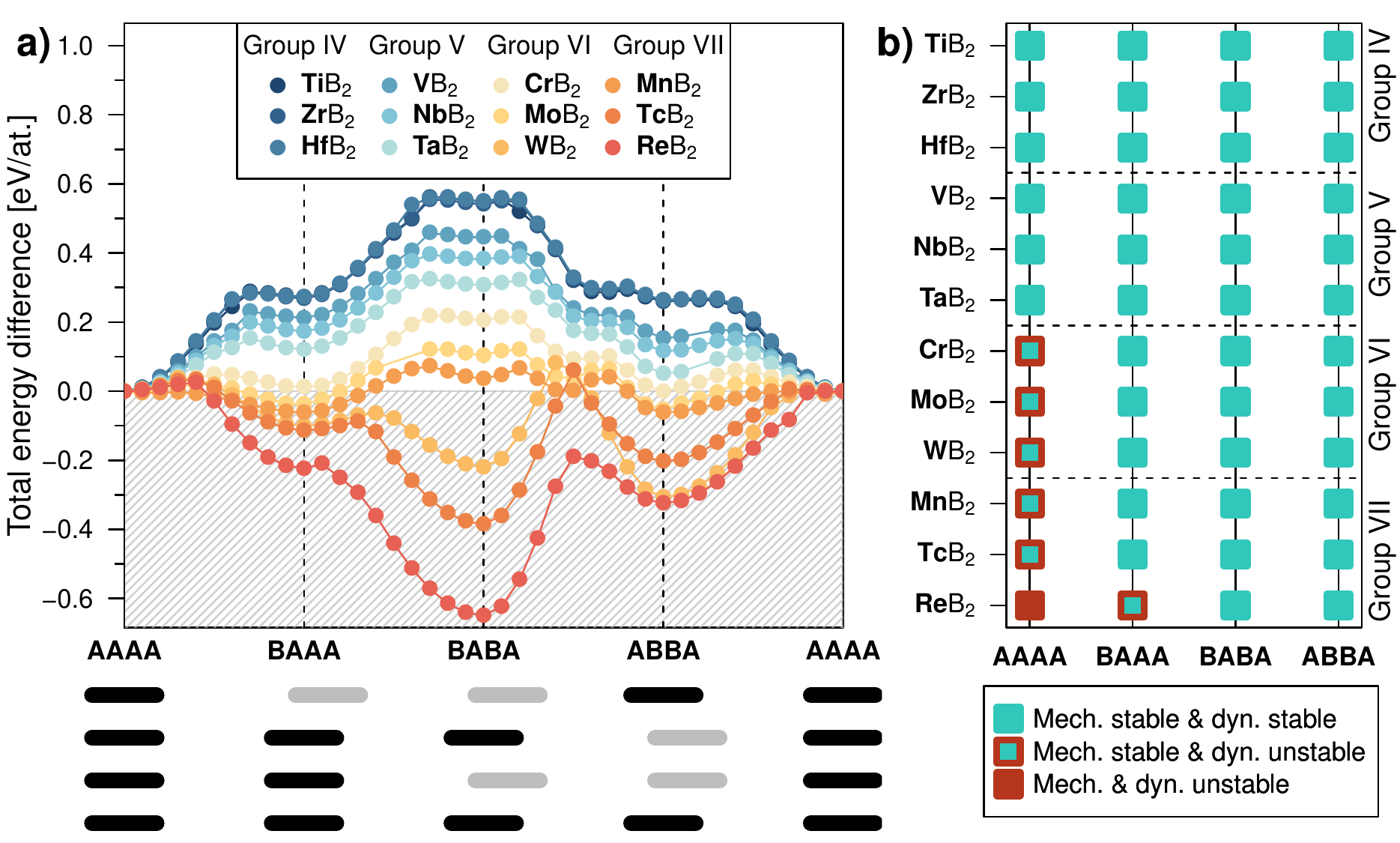} 
	\caption{
	(a) Total energy difference, $\Delta E_{\text{tot}}$, of TMB$_2$ structures (where TM are group IV--VII transition metals), with various stackings compared to the reference AAAA stacking. 
	The data points in-between stackings AAAA, BAAA, BABA, and ABBA correspond to step-wise shifts of the respective metallic plane(s) as schematically shown below the $x$-axis.  
	(b) Mechanical and dynamical stability of TMB$_2$ structures with AAAA, BAAA, BABA, and ABBA stacking sequences, respectively.
	}
\label{FIG: Fig1}
\end{figure*}

First we discuss total energy ($E_{\text{tot}}$) variations along the AAAA--BAAA--BABA--ABBA--AAAA transformation pathway (Fig.~\ref{FIG: Fig1}a), visualized in a relative comparison with the energy of the AAAA stacking,
\begin{equation}
\Delta E_{\text{tot}}=E_{\text{tot}}-E_{\text{tot}}(\text{AAAA})\ .  
\end{equation}
Groups IV and V TMB$_2$s yield positive $\Delta E_{\text{tot}}$ along the entire deformation path indicating that the AAAA stacking is the most stable one.
In contrast, virtually zero (see e.g. BAAA-CrB$_2$) or even largely negative $\Delta E_{\text{tot}}$ values (see e.g. BABA-ReB$_2$) calculated for the group VI and VII TMB$_2$s suggest comparable energetic preference or even a strong tendency for other stackings than the AAAA one.
This indication is further underpinned by the stability analysis (Fig.~\ref{FIG: Fig1}b) revealing that the AAAA allotrope is dynamically or even mechanically unstable (AAAA-ReB$_2$) for the group VI and VII TMB$_2$s.

Trends in the total energy variations (Fig.~\ref{FIG: Fig1}a) almost perfectly follow the left-to-right (group IV$\to$VII) and top-to-bottom (period 4$\to$6) move in the periodic table.
Specifically, $\Delta E_{\text{tot}}$ along the entire deformation pathway generally decreases when moving from the group IV (e.g. Ti) to VII (e.g. Re) transition metals---thus, when increasing the number of valence electrons---but also within each group when changing from the period 4 (e.g. V) to 6 (e.g. Ta)---thus, when increasing the number of electron shells. 
Starting with the group IV TMB$_2$s---TiB$_2$, ZrB$_2$, and HfB$_2$---Fig.~\ref{FIG: Fig1}a indicates a strong preference for the AAAA stacking, since both BAAA and ABBA yield $E_{\text{tot}}$ of about 0.28~eV/at. higher, and BABA shows $E_{\text{tot}}$ even $\approx{0.5}$~eV/at. above that of AAAA. 
The almost overlapping energy landscapes of TiB$_2$, ZrB$_2$, and HfB$_2$ suggest basically no effect of changing the period (Ti$\to$Zr$\to$Hf).
The group V TMB$_2$s---VB$_2$, NbB$_2$, and TaB$_2$---nearly mirror the $\Delta E_{\text{tot}}$ profile predicted for the group IV  TMB$_2$s, however, all energies are shifted to lower values.
Additionally, we observe a systematic $E_{\text{tot}}$ decrease for V$\to$Nb$\to$Ta, i.e. upon changing the period 4$\to$5$\to$6. 
While the BABA stacking is still the least stable one ($\Delta E_{\text{tot}}\approx{0.3}$--0.45~eV/at.), the energy profile becomes asymmetric due to a slight preference of the ABBA ($\Delta E_{\text{tot}}\approx{0.07}$--0.16~eV/at.) over BAAA ($\Delta E_{\text{tot}}\approx{0.16}$--0.21~eV/at.) stacking.
This is likely related to the fact that the former, ABBA, stacking is more symmetric than the BAAA one.
Despite relatively high energies of the BAAA, BABA, and ABBA stackings for the group IV and V transition metal diborides, they are all found dynamically stable (Fig.~\ref{FIG: Fig1}b). 

Moving to the group VI TMB$_2$s---CrB$_2$ (treated as non-magnetic), MoB$_2$, and WB$_2$---$\Delta E_{\text{tot}}$ shifts to even lower values, hence, changing the order of stability of the four allotropes (note that the AAAA stacking becomes dynamically unstable, cf. Fig.~\ref{FIG: Fig1}b).
With essentially zero $\Delta E_{\text{tot}}$ predicted for BAAA-CrB$_2$ and ABBA-CrB$_2$, the BAAA and ABBA stackings are energetically equivalent to AAAA, but are dynamically stable in contrast to the AAAA stacking.
Due to their negative $\Delta E_{\text{tot}}$ values (approx. $-0.04$ and $-0.05$~eV/at., respectively), the BAAA and ABBA variants of MoB$_2$ are even energetically preferred over the AAAA, and the ABBA stacking becomes the new lowest-energy allotrope.
With $\Delta E_{\text{tot}}\approx{-0.31}$~eV/at., the ABBA stacking is the most stable variant also for WB$_2$.
This is consistent with previous DFT calculations reporting that WB$_2$ prefers to crystallize in the $\omega$-type phase \cite{MoB2_WB2_structures}. 
As $\Delta E_{\text{tot}}$(BAAA-WB$_2$)$\approx{-0.10}$~eV/at. and $\Delta E_{\text{tot}}$(BABA-WB$_2$)$\approx{-0.22}$~eV/at., the BABA allotrope, which was the least stable one for the groups IV and V TMB$_2$, is more stable than both the AAAA and BAAA variants.

\add{
Since CrB$_2$ \cite{osti_1207357} and MnB$_2$ \cite{osti_1277089} have been reported as being ferromagnetic (in the AAAA structure), we have also calculated their total energies in the AAAA, BABA, and BBAA configurations. 
For all stackings, the total energy decreased with respect to the non-magnetic configurations.
More importantly, for CrB$_2$ the most stable allotrope changed from the AAAA to the ABBA stacking, whereas for MnB\textsubscript{$2$}, it changed from the ABBA to the BABA structure.
However, the magnetic degree of freedom adds huge complexity to the simulation protocol along the whole transformation path, which goes beyond the scope of the present overview study and hence will not be discussed anymore.
}

\add{
Fig.~\ref{FIG: Fig1}b summarizes the mechanical and dynamical stability of the investigated stacking.
The only mechanically unstable system is ReB$_2$ in the AAAA stacking, failing the condition $C_{44}>0$~\cite{mouhat2014necessary}.
This suggests that the AAAA-ReB$_2$ is unstable with respect to shear in the $y-z$ and $x-z$ planes.
And indeed, its stable configuration is the ABAB stacking (Fig.~\ref{FIG: Fig1}a): the local stacking change AA$\to$AB indeed corresponds to the out-of-plane shear $y-z$ or $x-z$ (or their combination).
}

The group VII TMB$_2$s---MnB$_2$ (treated as non-magnetic), TcB$_2$ (included for completeness but never experimentally reported), and ReB$_2$---yield mostly negative $\Delta E_{\text{tot}}$ values along the entire transformation pathway.
Similarly to the group VI TMB$_2$s, the AAAA allotrope is dynamically unstable. 
MnB$_2$ yields the lowest $\Delta E_{\text{tot}}$ ($\approx{-0.06}$\;eV/at.) for the ABBA stacking, closely followed by BAAA.
TcB$_2$ and ReB$_2$ exhibit very deep global energy minima at the BABA stacking, with 
$\Delta E_{\text{tot}}$ of $-0.38$ and $-0.65$~eV/at., respectively.
This agrees well with the previously reported preference of ReB$_2$ for the $\gamma$-type phase \cite{ReB2_WB2_paper}. 
Also the BAAA and ABBA allotropes of ReB$_2$ show low $\Delta E_{\text{tot}}$ values, both below $-0.2$\;eV/at., however, the former is predicted to be dynamically unstable.

\begin{figure*}[h!t!]
	\centering    
    \includegraphics[width=1.75\columnwidth]{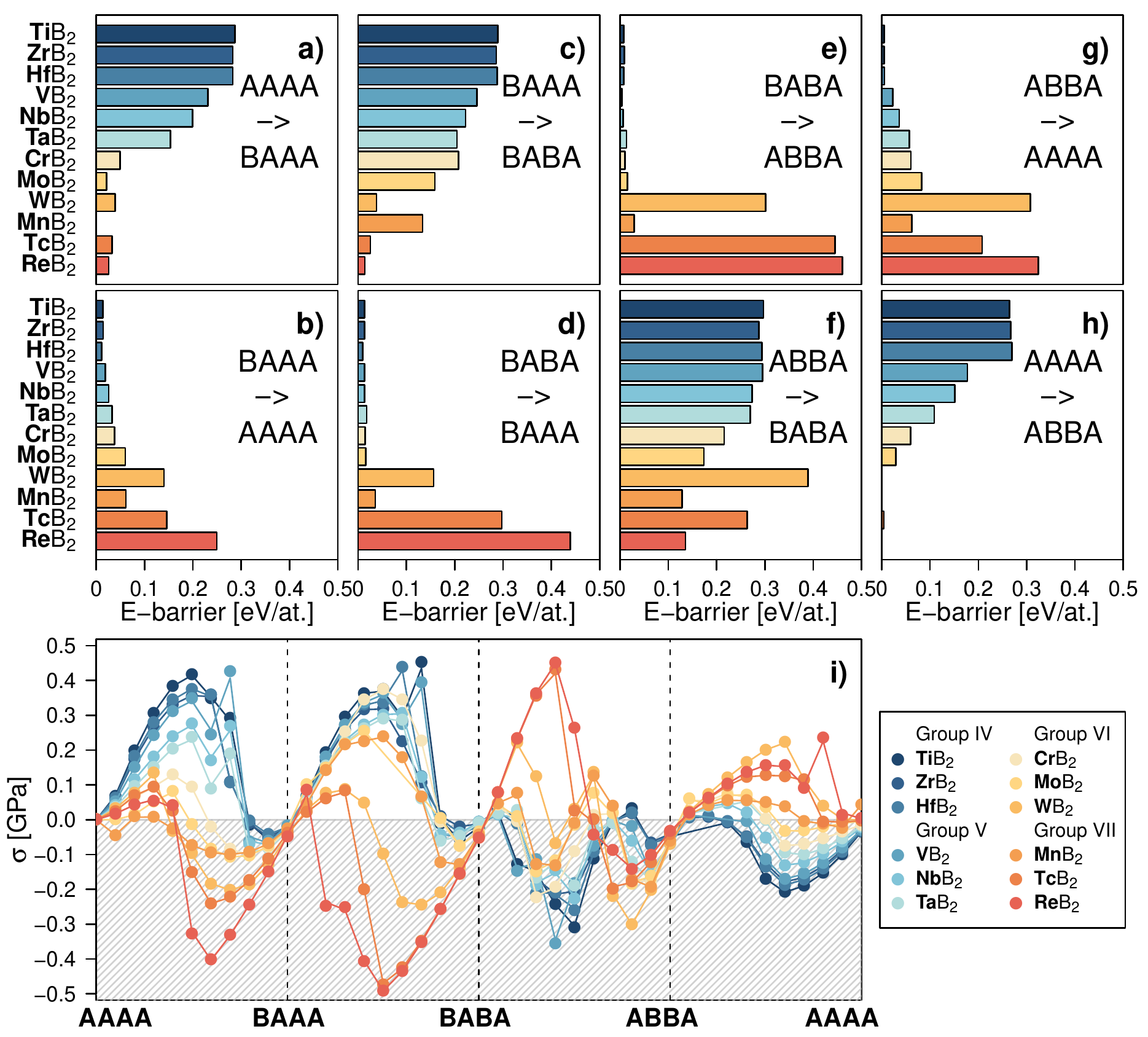}
	\caption{
	Energy barriers (defined by Eq.~\ref{EQ: E-barrier}) for the (a) AAAA$\to$BAAA, (b) BAAA$\to$BAAA, (c) BAAA$\to$BABA, (d) BABA$\to$BAAA, (e) BABA$\to$ABBA, (f) ABBA$\to$BABA, (g) ABBA$\to$AAAA, (h) AAAA$\to$ABBA transitions.
	(i) The normal stress, $\sigma_{xx}$ (defined by Eq.~\ref{Eq: sigma_xx}), along the $\langle1\bar100\rangle$ direction plotted in relative comparison to the AAAA stacking.
	}
\label{FIG: Fig2}
\end{figure*}

Fig.~\ref{FIG: Fig2}a--h depict energy barriers, $E$-barriers (for definition, please see Eq.~\eqref{EQ: E-barrier} in the Methods), that need to be overcome when changing between the AAAA, BAAA, BABA, and ABBA diboride allotropes. 

The AAAA$\to$BAAA transformation (Fig.~\ref{FIG: Fig2}a) comes with high energetic costs (0.28--0.29\;eV/at.) for the group IV TMB$_2$, decreasing for the group V TMB$_2$ (0.23--0.15\;eV/at. for V$\to$Nb$\to$TaB$_2$), and almost diminishing for the group VI and VII TMB$_2$, which fall down to 0--0.05\;eV/at.
Conversely, the BAAA$\to$AAAA transition (Fig.~\ref{FIG: Fig2}b) is energetically cheap for the group IV and V TMB$_2$ (0.01--0.03\;eV/at.) and becomes more costly for their group VI--VII counterparts, with the highest barrier of $\approx{0.25}$\;eV/at. predicted for ReB$_2$. 

The barrier associated with the BAAA$\to$BABA transition (Fig.~\ref{FIG: Fig2}c) is again the highest for the group IV TMB$_2$ (0.29\;eV/at.) and decreases down to 0.01\;eV/at. when moving to the right in the periodic table, i.e. to the group VII TMB$_2$. 
It also decreases within each group (with MnB$_2$ being the only outlier from this trend). 
Transformation in the opposite direction, BABA$\to$BAAA (Fig.~\ref{FIG: Fig2}d), presents basically no energetic barrier (0.01--0.02\;eV/at.) for the group IV--VI TMB$_2$---excluding WB$_2$ with $E$-barrier of 0.16\;eV/at.---while the same transition becomes very costly for TcB$_2$ and ReB$_2$ (with $E$-barrier above 0.3\;eV/at.). 

For the group IV--V \add{T}MB$_2$ together with CrB$_2$ and MoB$_2$, the BABA$\to$ABBA transition (Fig.~\ref{FIG: Fig2}e) requires the energy of only 0.01--0.02\;eV/at., while the reverse ABBA$\to$BABA is associated with a barrier of 0.17--0.3\;eV/at.
Both the BABA$\to$ABBA and ABBA$\to$BABA transitions are relatively costly (above 0.13\;eV/at.) for TcB$_2$ and ReB$_2$. 

Finally, the ABBA$\to$AAAA and AAAA$\to$ABBA (Fig.~\ref{FIG: Fig2}g--h) energy barriers for the group IV TMB$_2$ are 0.01 and 0.27\;eV/at., respectively, underpinning the strong preference for AAAA stacking.  
The ABBA$\to$AAAA transition is cheap also for the group V TMB$_2$ ($E$-barrier of 0.02--0.06\;eV/at.), while the AAAA$\to$ABBA again comes at higher energetic costs (0.18--0.11\;eV/at.).
CrB$_2$ presents the borderline with nearly the same barriers (0.06\;eV/at.) for the ABBA$\to$AAAA and AAAA$\to$ABBA transformation.
Afterward going to MoB$_2$, WB$_2$, MnB$_2$, TcB$_2$, and ReB$_2$, the AAAA$\to$ABBA transition is associated with much lower barrier compared to the ABBA$\to$AAAA transition.

We note that since the AAAA stacking is dynamically unstable for the group VI--VII TMB$_2$, barriers involving the AAAA structure of those systems should be taken with a grain of salt; instead, we propose that a direct transition BAAA$\leftrightarrow$BBAA (which is due to periodic boundary conditions equivalent with the ABBA stacking) will take place, presumably yielding lower barriers than the ABBA$\leftrightarrow$AAAA$\leftrightarrow$BAAA ones.

To interpret the transformation energetics in terms of mechanical loading, we considered the supercell as a solid box composed of four blocks. 
During each transformation step, one or two of these blocks ``move'' (cf. the scheme in Fig.~\ref{FIG: Fig1}a).
The motion of a block $i$ can be viewed as a consequence of an applied horizontal force $F_i$ acting along the $\langle1\bar100\rangle$ direction.
When a block $i$ changes its stacking type during the transformation step, i.e. $A\to B$ or $B\to A$, the force $F_i$ facilitates work $W_i$ along the path of a length $|\langle1\bar100\rangle|a/3 = \sqrt{3}a/3$, that is
\begin{equation}
    W_i(x) = \int_0^x F_i(\xi) \frac{\sqrt{3}a}{3} d\xi\ .
\end{equation}
If the stacking does not change, the path has a zero length and no work is done.
The total energy change along the transformation path (Eq.~\ref{EQ: E-barrier}) equates to a sum of the $W_i$ contributions, i.e. the work done starting from the initial AAAA configuration,
\begin{equation}
    \Delta E_{\text{tot}}(x) = \sum_{i=1}^4 W_i(x)\ .
\end{equation}
Considering the mirror symmetry of the simulation box, the instantaneous magnitudes of the acting forces (i.e. acting on moving blocks) are the same. 
Therefore, the force magnitude can be obtained as a derivative of the energy
\begin{equation}
    F(x) = \frac{1}{n}\cdot \frac3{\sqrt3a}\cdot \frac{d }{dx} \bigg( \Delta E_{\text{tot}}(x)\bigg)\ ,
    \label{Eq: F(x)}
\end{equation}
where the multiplicity factor $n$ (the number of moving blocks) is 1 along the AAAA$\leftrightarrow$BAAA$\leftrightarrow$BABA pathways, and 2 otherwise.
Finally, we recalculate the force $F(x)$ (Eq.~\ref{Eq: F(x)}), to the applied normal stress, $\sigma$ in the $\langle1\bar100\rangle$ direction by dividing $F(x)$ by the normal area $A$ of the block, 
\begin{equation}
    \sigma(x) = \frac{F(x)}{A} = F(x)\cdot\frac{4}{ac}\ ,
    \label{Eq: sigma_xx}
\end{equation}
where $c$ is the simulation box length along the $[0001]$ direction for the reference AAAA stacking.
The resulting $\sigma(x)$ profile is shown in Fig.~\ref{FIG: Fig2}i, indicating that the maximum stresses along the AAAA$\to$BAAA$\to$BABA path are larger than for AAAA$\to$ABBA$\to$BABA transformation path for all group \add{I}V and V TMB$_2$, as well as for CrB$_2$, MoB$_2$ and MnB$_2$.
This suggests that the $\alpha\to\gamma$ transformation---when facilitated by normal stresses and related shuffling of planes---may proceed differently for different TMB$_2$.
Unlike that, the $\alpha\to\omega$ transformation is predicted to always proceed directly AAAA$\to$ABBA, rather than via the \add{$\gamma$}\ BABA stacking.
The lowest transformation stress is predicted for ReB$_2$: 0.09\,GPa for the $\alpha\to\gamma$ and 0.03\,GPa for $\alpha\to\omega$ transformations. However, the most stable configuration of the ReB$_2$ is the BABA stacking; the transformation stresses from $\gamma\to\alpha$ and $\gamma\to\omega$ reach over 0.4\,GPa, thus making this allotrope extremely stable, once formed.
On the contrary, TiB$_2$ yields the largest transformation stress for the $\alpha\to\omega$ (0.21\,GPa), whereas for the $\alpha\to\gamma$ transformation, the maximum stress is obtained for VB$_2$ (0.36\,GPa) proceeding via the ABBA ($\omega$) stacking.

\begin{figure*}[h!t!]
	\centering
    \includegraphics[width=2\columnwidth]{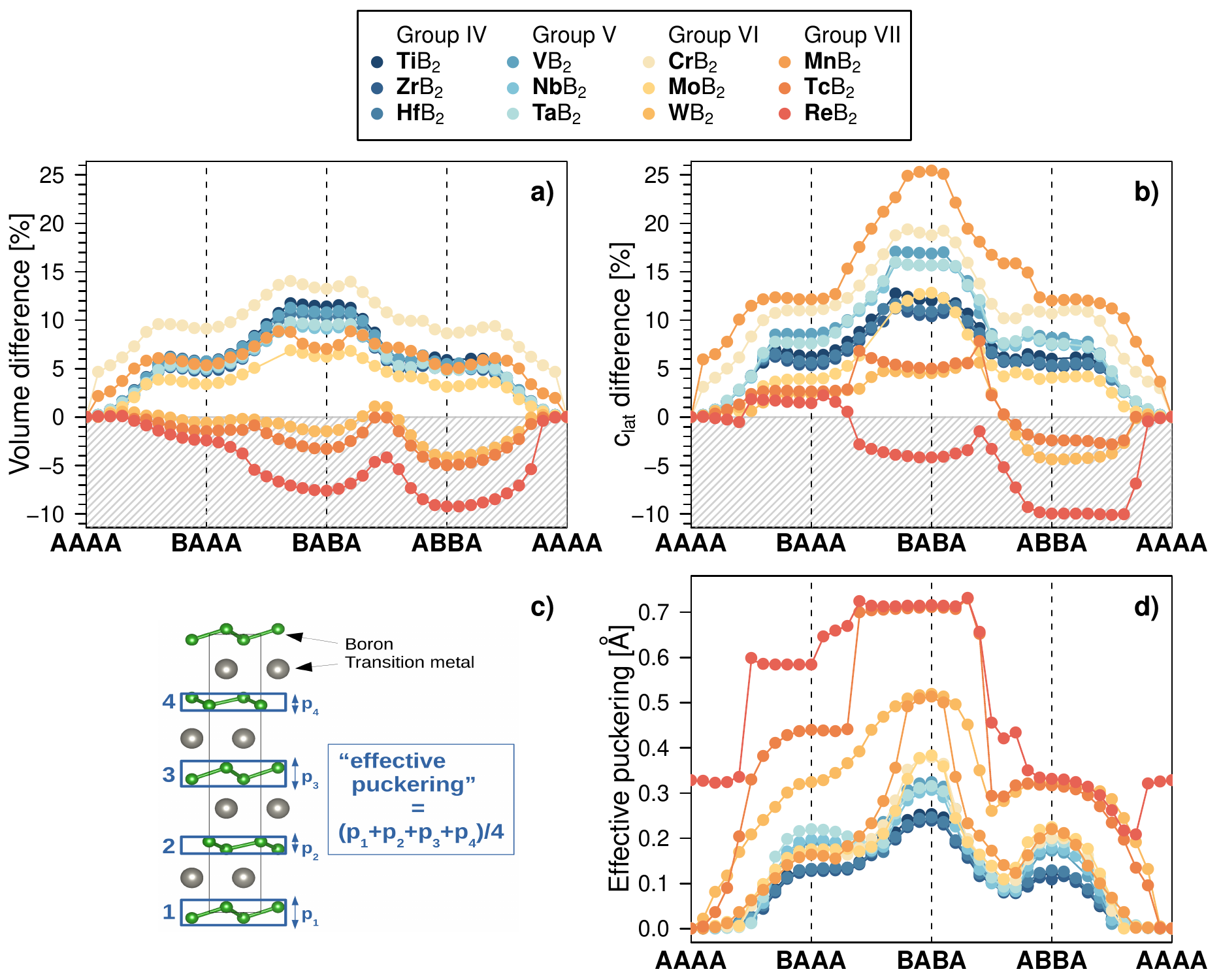}
	\caption{
	(a) Volume and (b) $c$ lattice parameter variations along the AAAA--BAAA--BABA--ABBA--AAAA transformation pathway.
	Positive (negative) values [\%] denote relative increase (decrease) compared to the reference AAAA stacking.
	(c)~Schematic definition of the {\it{effective puckering}} of boron planes, depicted in panel (d).
	}
\label{FIG: Fig3}
\end{figure*}

\subsection{Structural changes}
Energy variations along the AAAA--BAAA--BABA--ABBA--AAAA transformation pathway can be further understood in view of the underlying structural changes (Fig.~\ref{FIG: Fig3}).
In particular, different stacking sequences on the metal sublattice are followed by (partial) puckering of the boron hexagons and, consequently, volumetric changes.

Fig.~\ref{FIG: Fig3}a depicts relative volume ($V$) increase/decrease in comparison with the AAAA stacking.
The group IV--V transition metal diborides yield the overall lowest volume for the AAAA stacking, earlier identified as their lowest-energy allotrope (Fig.~\ref{FIG: Fig1}a). 
Furthermore, these diborides exhibit a volume increase along the entire deformation pathway, with maximum volume (i.e. the lowest density) predicted for the energetically least favorable BABA stacking. 
For illustration, the AAAA$\to$BAAA and AAAA$\to$ABBA TiB$_2$ transitions lead to $\approx{5}$\% volume increase, while the AAAA$\to$BABA TiB$_2$ transition enlarges volume by $\approx{11}$\%.
The AAAA stacking has the lowest volume also for CrB$_2$, MoB$_2$, and MnB$_2$, however, we recall its dynamical instability according to Fig.~\ref{FIG: Fig1}b.
CrB$_2$ also shows the overall greatest volume increase along the entire pathway which, however, might stem from omitting its magnetism.
Worth highlighting is the volume decrease predicted for the BAAA, BABA, and ABBA allotropes of WB$_2$, TcB$_2$, and ReB$_2$.  
We recall dynamical instability of the AAAA allotrope according to Fig.~\ref{FIG: Fig1}b.
Specifically for BABA-ReB$_2$, the lowest-energy ReB$_2$ allotrope, Fig.~\ref{FIG: Fig3}a reveals a volume decrease by $\approx{8}$\% compared to AAAA-ReB$_2$.

Volume changes in Fig.~\ref{FIG: Fig3}a mainly stem from the evolution of the $c$ lattice parameter Fig.~\ref{FIG: Fig3}b, which for almost all diborides---with the exception of WB$_2$, TcB$_2$, and ReB$_2$---increases when leaving the perfect AAAA stacking. 
The lattice parameter $c$ is parallel to the hexagonal [0001] direction, thus orthogonal to the metal/boron layers.  
The increase in $c$ is compensated by relatively small lateral shrinkage (by up to 4\%), i.e. lattice parameter $a$ ($=b$) decrease (not shown).

Furthermore, we investigate how the different stackings of the metal planes influence the boron sublattice, in particular, the puckering of boron hexagons. 
Fig.~\ref{FIG: Fig3}c shows a schematic definition of the {\it{effective puckering}} ($p_{\text{eff}}$), calculated using the thicknesses of four boron planes in our simulation cell.
Similarly to $\Delta E_{\text{tot}}$, $p_{\text{eff}}$ shows a strong trend following the left-to-right (group IV--VII TMB$_2$) and top-to-bottom (period 4--6 TMB$_2$) shift in the periodic table. 
Starting with the group IV TMB$_2$---TiB$_2$, ZrB$_2$, and HfB$_2$---the $p_{\text{eff}}$ evolution shows essentially the same profile as $\Delta E_{\text{tot}}$, $V$, and $c$: with similar values for the BAAA and ABBA stackings, and a peak at the least energetically stable BABA.
Changing to the group V--VII TMB$_2$, boron layers gradually pucker more significantly, compare, e.g. TiB$_2$, TaB$_2$, and WB$_2$.
Such pronounced puckering is for most diborides mirrored by $c$ lattice parameter and volume increase.
Interestingly, boron plane puckering in WB$_2$, TcB$_2$, and ReB$_2$ is associated with a volume decrease.
Furthermore, $p_{\text{eff}}$ in the above three diborides significantly increases immediately after leaving the perfect AAAA stacking, which is consistent with high metastability/instability of the AAAA allotrope for these TMB$_2$.

\begin{figure*}[h!t!]
	\centering
    \includegraphics[width=2\columnwidth]{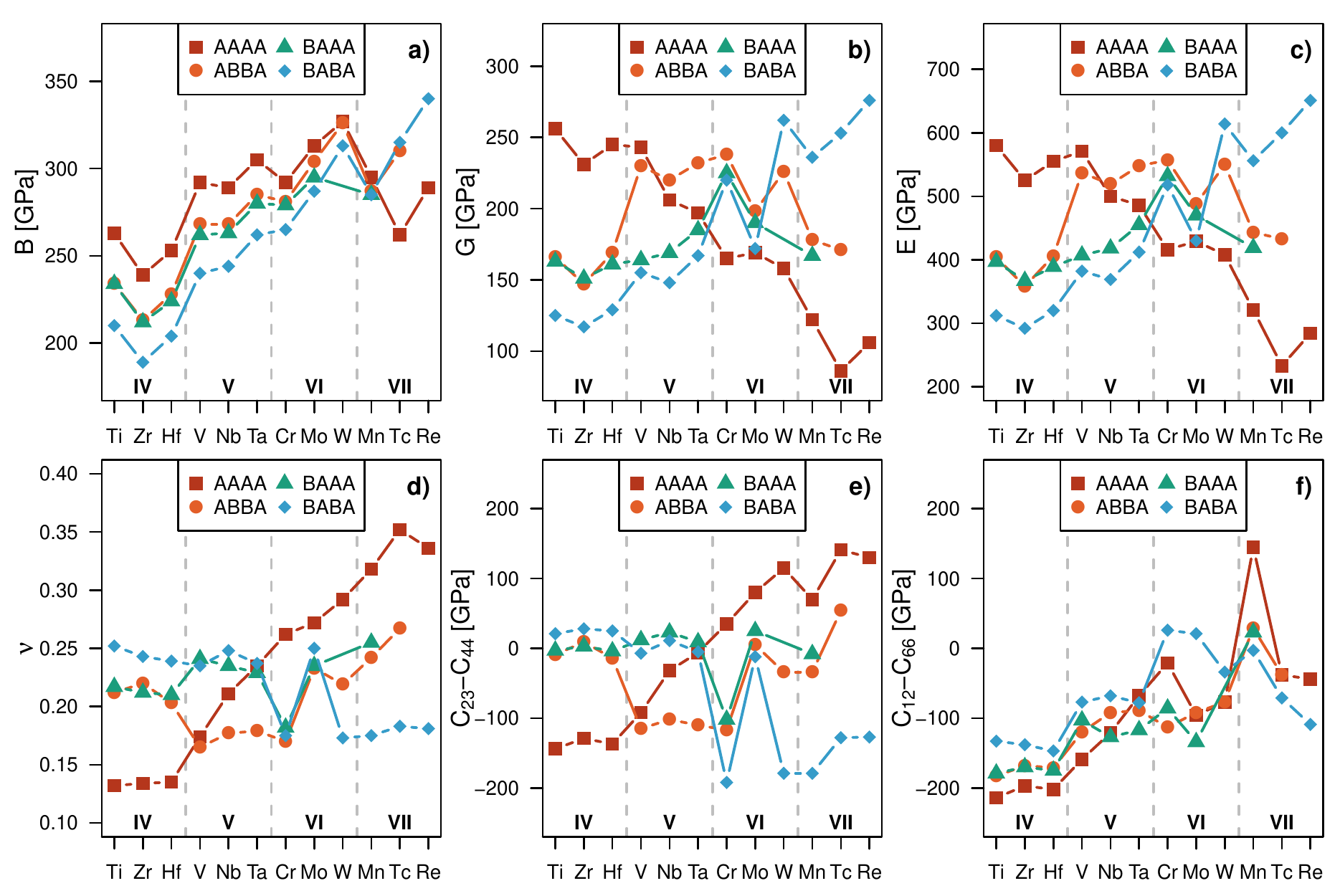}
	\caption{
	Trends in mechanical properties of $M$B$_2$ structures (where $M$ are group IV--VII transition metals) with AAAA, BAAA, ABBA, and BABA stacking sequences, as estimated based on elastic constants ($C_{ij}$) calculations.
	Polycrystalline (a) bulk, (b) shear, and (c) Young's moduli, together with ductility approximates: (d) Poisson's ratio, and Cauchy pressure, (e) $CP_1=C_{23}-C_{44}$ and (f) $CP_2=C_{12}-C_{66}$. Higher values indicate more metallic bonding, thus, increased tendency for ductile behavior.  
	}
\label{FIG: Fig4}
\end{figure*}

\subsection{Mechanical properties}
Transitions between the AAAA, BAAA, ABBA, and BABA allotropes have consequences also for mechanical properties, which can be estimated via elastic constants, $C_{ij}$.  
In Fig.~\ref{FIG: Fig4}, we plot trends in polycrystalline bulk ($B$), shear ($G$), and Young's moduli ($E$), together with ductility estimates based on the Poisson's ratio ($\nu$), and Cauchy pressure $C_{23}-C_{44}$ and $C_{12}-C_{66}$.
Note that mechanically unstable systems (based on the criteria for elastic constants in Ref.~\cite{mouhat2014necessary}) are not shown.
For the group IV--VI TMB$_2$s, trends in the bulk modulus (Fig.~\ref{FIG: Fig4}a) seem {to resemble} the energetic stability trends (Fig.~\ref{FIG: Fig1}a) in a way that the highest $B$ is shown by the energetically most stable AAAA stacking, the lowest $B$ for the least stable BABA stacking, and the ABBA and BAAA stackings---energetically in-between AAAA and BABA---exhibit $B$ values between those of the AAAA and BABA stacking.
Furthermore, as the energetic differences between the four allotropes diminish when going from group IV to VI, so do differences in their bulk moduli.
While we do not see any clear explanation for the similarity between the total energy and bulk modulus trends, it could relate to volumetric changes in Fig.~\ref{FIG: Fig3}a, where the relative volume increase with respect to the AAAA stacking (followed by an energy increase) could induce lower resistance to compression.
The overall highest bulk modulus ($\approx{340}$\;GPa) is predicted for the BABA-ReB$_2$, i.e. the lowest-energy ReB$_2$ allotrope (with $\approx{8}$\% lower volume compared to the AAAA stacking), followed by the ABBA-WB$_2$, i.e. the lowest-energy WB$_2$ allotrope (with $\approx{5}$\% lower volume compared to the AAAA stacking).
Please also recall that AAAA-ReB$_2$ and AAAA-WB$_2$ are dynamically unstable.
The shear and Young's moduli (Fig.~\ref{FIG: Fig4}b,c) evolve in a similar manner, differing from the relatively simple trend predicted for the bulk modulus. 
In particular, $G$ ($E$) of the AAAA stacking decreases from {$G=256$\;GPa} (Ti) to {$G=106$\;GPa} (Re) (from {$E=580$}\;GPa to {$E=284$\;GPa}) when moving from the group IV to VII TMB$_2$s.
In contrast, $G$ ($E$) of the BABA stacking increases from {$G=125$\;GPa} (Ti) to {$G=276$\;GPa} (Re) (from {$E=312$}\;GPa to {$E=651$\;GPa}) when moving from the group IV to VII TMB$_2$s.
Shear and Young's moduli of the ABBA and BAAA stackings show \add{a} relatively lower spread, nonetheless, increase for the group V--VI TMB$_2$s for which these two stackings are associated with low or almost zero energy barriers. 
Similar to the bulk modulus, the overall highest $G$ and $E$ values are predicted for BABA-ReB$_2$, pointing towards superior strength of this material.
This is consistent with literature reports claiming ultra-incompressibility and superhardness of ReB$_2$ \cite{levine2010full,ultra_high_hardness_ReB2}.
Our $B$ and $G$ moduli yield a good agreement with {\it{ab initio}} calculated values for $\alpha$-structured diborides of the group IV--VI transition metals predicted by \citet{gu2021sorting}.

As other ceramics, transition metal diborides are hard but suffer from brittleness which is a strong limiting factor for their fracture toughness \cite{magnuson2021review}. 
Within the same material class, Poisson's ratio and Cauchy pressure are widely accepted empirical indicators allowing to compare two or more systems in terms of their metallic/covalent bonding character, providing a basis for more ductile/brittle behavior \cite{greaves2011poisson,pettifor1992theoretical}.  
Examples within the transition metal nitride family include Refs.~\cite{koutna2021high,sangiovanni2010electronic,balasubramanian2018valence}.
The Poisson's ratio (Fig.~\ref{FIG: Fig4}d) of the AAAA allotrope is the lowest($\approx{0.13}$) for the group IV-TMB$_2$---TiB$_2$, ZrB$_2$, HfB$_2$---and significantly increases (up to $\approx{0.35}$) when moving to the group V, VI, and VII TMB$_2$s, hence, suggesting improved ductility with increased valence electron concentration (VEC).
An inverse (decreasing) trend is predicted for the BABA allotrope, while the ABBA and BAAA allotropes show nearly overlapping $\nu$ values for the group IV, VI, and VII TMB$_2$ but differ for the group V TMB$_2$s.
Focusing only on the lowest-energy stacking of each element, our calculations indicate similar brittleness/low ductility of TiB$_2$, ZrB$_2$, and HfB$_2$, which significantly improves when going VB$_2$, NbB$_2$, and TaB$_2$.
Changing to CrB$_2$, ductility indicators drop again, while those of MoB$_2$ and WB$_2$ are comparable to that of TaB$_2$.
The lowest-energy allotropes for the group VII-TMB$_2$ are predicted to be comparably brittle/moderately ductile as VB$_2$. 
We note that the here predicted $\nu$ values for $\alpha$- and $\omega$-structured diborides agree well with {\it{ab initio}} calculations by \citet{moraes2018ab}.
In contrast to Poisson's ratio, Cauchy pressure (Fig.~\ref{FIG: Fig4}e--f) can provide \add{a} directionally-resolved indication of ductility.
We recall that \add{the} Cauchy pressure of material with cubic symmetry is defined as $CP=C_{12}-C_{44}$.
Since for hexagonal structures $C_{12}\neq C_{13}=C_{23}$ and $C_{44}=C_{55}\neq C_{66}$, one can define $CP_{1}=C_{23}-C_{44}=C_{13}-C_{55}$ and $CP_{2}=C_{12}-C_{66}$.
The predicted higher $CP_{1}$ values---compared to $CP_{2}$---therefore indicate relatively more ductile (less brittle) character of the (more widely spaced) basal planes compared to the prismatic planes.

\begin{figure}[h!t!]
	\centering
    \includegraphics[width=\columnwidth]{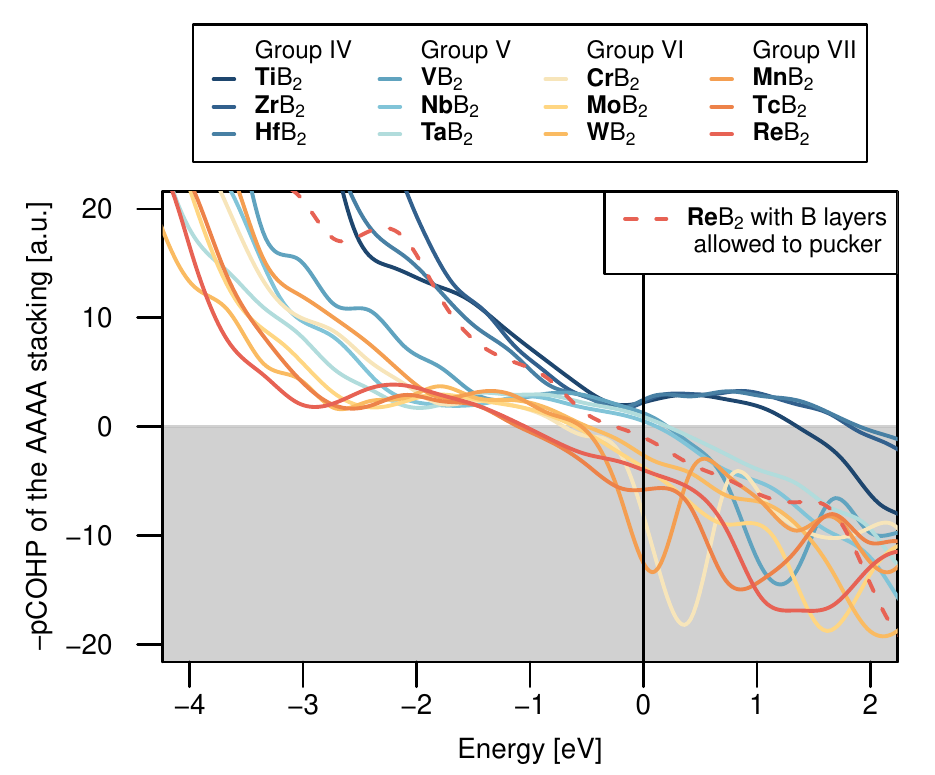}
	\caption{\add{Average projected COHP calculated for AAAA stacking. Negative values (gray region) correspond to destabilizing contributions.}
	}
\label{FIG: Fig5}
\end{figure}

\begin{figure*}[h!t!]
	\centering
    \includegraphics[width=2\columnwidth]{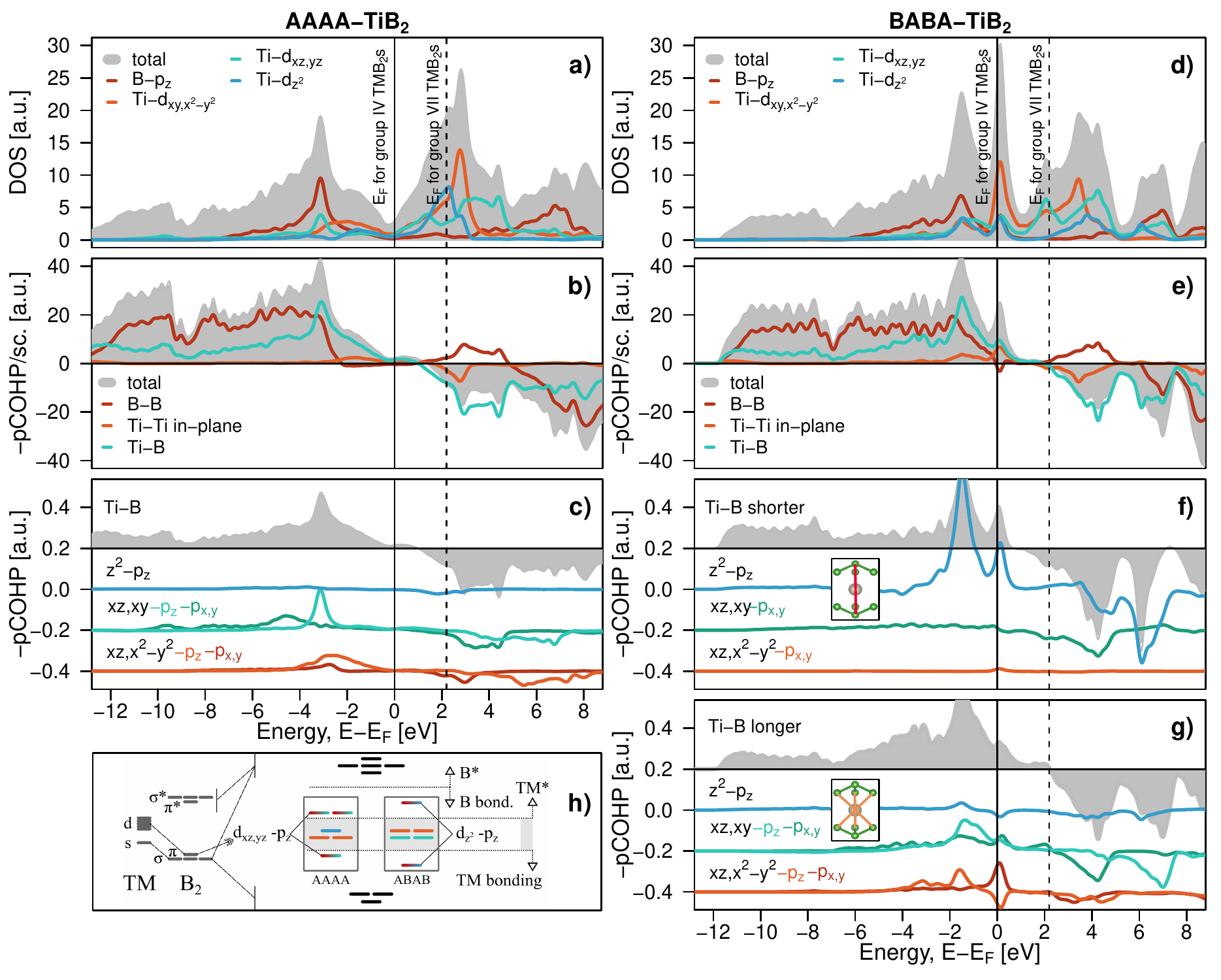}
	\caption{\add{For TiB$_2$ in AAAA structure calculated (a) DOS, (b) total projected COHP, (c) TM-B (2.38\AA) orbital resolved projected COHP/bond. Calculations in BABA structure: (d) DOS, (e) total projected COHP and orbital resolved projected COHP/bond for (f) shorter TM-B (2.24\AA) and (g) longer TM-B (2.42\AA). (h) Suggested MO diagram for TMB$_2$. The gray area for non-bonding TM levels, the asterisk denotes antibonding states. Colors of $d$ states match local DOS in panels a,d..}
	}
\label{FIG: Fig6}
\end{figure*}


\add{\subsection{Electronic structure analysis}}
\add{The collection of average COHP curves for all TMB\textsubscript{$2$}s in the AAAA structure summarized in Fig. \ref{FIG: Fig5}, shows trends consistent with our calculated transformation paths (Fig. \ref{FIG: Fig1}). The stability of the AAAA stacking gradually decreases as we move to higher group TMs. This fact can be explained by the occupation of antibonding (destabilizing) states as we increase the valence electron count. The COHP curves of group IV diborides contain only bonding states below the Fermi levels ($E_F$) while the presence of the antibonding states increases as we proceed to group VII. The preference for the AAAA stacking, therefore, downgrades in higher group TMB\textsubscript{$2$}s---an observation reflected by the diminishing energy barriers between AAAA and other allotropes (see Fig. \ref{FIG: Fig2}a--h). Additionally, when looking at ReB\textsubscript{$2$}, one can notice a reduction of the antibonding states below the Fermi level in the structure, including puckering of the boron sheets (the AAAA stacking in ReB\textsubscript{$2$} prefers to pucker and translate every second layer of boron, dashed line in Fig. \ref{FIG: Fig5}).} 

\add{To elaborate on the idea of the destabilizing effect of the antibonding states, in Fig. \ref{FIG: Fig6} we compare TiB\textsubscript{$2$} in the AAAA ($\alpha$) and BABA ($\gamma$) structures. By looking at the COHP curves we find two strong interactions, namely B-B that is bonding even above $E_F$ and TM-B that becomes destabilizing at lower energies due to the occupation of the antibonding metal states \cite{chen2008electronic, wang2019designing}.}

\add{The main differences between both structures include redistribution of $d_{z^2}$ and in-plane $d_{xy, x^2-y^2}$ orbitals. Firstly, when going from AAAA to BABA, the $d_{z^2}$ DOS peak (blue line in Fig. \ref{FIG: Fig6}a and d)  that is non-bonding in AAAA splits due to an interaction with B $p_z$ (red line in Fig. \ref{FIG: Fig6}a and d). Secondly, localization of the in-plane $d$ orbitals (orange line in Fig. \ref{FIG: Fig6}a and d) is enhanced in BABA due to a stronger in-plane interaction between metal atoms induced by the TM-TM bond shortening. Even though bonding in nature, the localized TM-TM peak in BABA shifts to the pseudo-gap, which results in considerably increased DOS at $E_F$, making the BABA structure unfavorable for early TMB\textsubscript{$2$}s. To reason the stabilization of BABA in higher group TM diborides, the role of $d_{z^2}$ orbitals was proposed to be crucial \cite{burdett1986electronic}. Within the AAAA structure, metal atoms
positioned below the center of boron hexagons form antibonding TM-TM interaction across the boron sheets. This unfavorable interaction can be relieved by the translation of the metal atoms at positions directly below the B atoms, which also reduces the TM-B coordination from 12 to 8. Further stabilization is achieved by puckering of boron hexagons \cite{burdett1986electronic}. Albeit the reduction of the antibonding TM-TM overlap of the $d_{z^2}$ orbitals has a stabilizing character, the energetic contribution of this bond in negligible compared to B-B and TM-B. A charge transfer from TM to B allows for the formation of strong $\pi$ bonds within the boron hexagons making the stabilizing contribution of B-B bonds greater than that of TM-B bonds.}

\add{As we move to the higher group TMs, the charge transfer reduces, and $d$ orbitals fill up, which strengthens the TM-B bond. Consequently, TM-B bonds in early TMB\textsubscript{$2$}s are longer, causing larger volume and narrower puckering of the boron sheets along the $c$-axis (compare with Fig. \ref{FIG: Fig3}). The interplay between the two strongest bonds is, therefore, one of the leading parameters in the stability of TMB\textsubscript{$2$}s.}

\add{Taking advantage of the rigid band approach \cite{chen2008electronic, wang2019designing, burdett1986electronic}, we can generalize the TiB\textsubscript{$2$} results to other TMB\textsubscript{$2$}s. In Fig. \ref{FIG: Fig6}a--g the dashed lines denote E\textsubscript{$F$} in group IV and VII diborides. Focusing on group VII TMB\textsubscript{$2$}s, the destabilization of the AAAA structure can be related to the filling of the DOS peak above the pseudo-gap with mostly metallic character that comes hand in hand with filling of the antibonding states \cite{chen2008electronic}, as visible in Fig. \ref{FIG: Fig6}a--b. On the contrary, within the BABA structure the bonding/antibonding transition is shifted to higher energies.}

\add{To explain this behavior, we propose a simple molecular orbital model (Fig. \ref{FIG: Fig6}h). First we assume the $sp^2$ hybridization of B atoms that consequently interact with TM atoms (grey levels on the left in Fig. \ref{FIG: Fig6}h). The metal $s$ electrons then interact with the $\sigma$ orbitals to form the low-lying bonding region and, together with antibonding $sp^2$, the high energy antibonding region (black levels common to the AAAA and BABA structures). Next, the B $\pi$ electrons, which are mostly of the $p_z$ character, hybridize with TM $d$ electrons. In the case of AAAA B $p_z$ predominantly interacts with the out-of-plane $d_{xz, yz}$ forming the distinct covalent peak at $-3$\,eV in TiB\textsubscript{$2$} (see also the orbital resolved COHP in the panel c).  This interaction leaves $d_{z^2}$ and in-plane $d_{xy,x^2-y^2}$ orbitals nonbonding (the shaded region between TM bonding/antibonding states in Fig. \ref{FIG: Fig6}h). On the other hand, in the BABA structure TM planes move and alter the stacking such that they reside at  in-line positions with B atoms. This composition is favorable for a strong $d_{z^2}$-$p_z$ interaction ($c$-axis in-line TM-B interaction in Fig. \ref{FIG: Fig6}f) leaving both out and in-plane $d$ orbitals non-bonding, as sketched in panel h.}

\add{In real extended systems, in-plane TM-TM interactions that were neglected in the simplistic orbital model split $d$ orbitals into bonding and antibonding levels Fig. \ref{FIG: Fig6}c and g). However, the position of metal atoms below the center of B hexagons in AAAA leaves the $d_{z^2}$ levels non-bonding Fig. \ref{FIG: Fig6}c). The metal-metal interaction in AAAA, therefore, splits only the in-plane $d_{xy, x^2-y^2}$ orbitals contrary to the splitting of both in and out-of-plane orbitals in BABA. This effectively creates more available bonding states in BABA and shifts the bonding/antibonding turning point to higher energies. Hence filling of the $d$ orbitals brings about stacking alternation to form short TM-B bonds and stabilizes structures with the puckered boron layers.    }
\\
\section{Conclusions}
We used first-principles calculations to simulate transformations between the well-known $\alpha$ (P6/mmm), $\gamma$ (P6$_3$/mmc), and $\omega$ (P6$_6$/mmc) phase prototypes of transition metal diborides (TMB$_2$s).
Alternating purely B and purely TM layers, the prototypes were regarded as different stackings of the TM planes, where the AAAA, BABA, and ABBA stackings correspond to the $\alpha$, $\gamma$, and $\omega$ structure, respectively.
Subsequently, transformations along the AAAA$\to$BAAA$\to$BABA$\to$ABBA pathway were facilitated by sliding of TM layers.
We discussed the predicted chemistry-related trends for the group IV--VII TMB$_2$s, focusing on energetics, stability, structural changes, and changes in elastic properties.

Total energy variations along the transformation pathway decrease when going from group IV to V TMB$_2$s, for which all stackings are found dynamically stable, with the following order of stability: AAAA$<$BAAA$\sim$ABBA$<$BABA.
The energy barriers are rather small (below 0.03\;eV/at.) when moving towards the $\alpha$-structure, but comparatively high (up to 0.29\;eV/at.) when moving away from it, suggesting that the metastable phases easily transform to the lowest-energy $\alpha$ phase.
The AAAA, BAAA, BABA, and ABBA allotrope change their order of stability for the group VI TMB$_2$s and, moreover, the AAAA stacking becomes dynamically unstable.
The ABBA variant is predicted to be the most stable phase prototype for both MoB$_2$ and WB$_2$.
For the group VII TMB$_2$s, the AAAA allotrope remains dynamically unstable.
While MnB$_2$ energetically prefers the ABBA stackings, it exhibits low energy barriers (around 0.1\;eV/at.) for the other two dynamically stable allotropes.
Unlike that, TcB$_2$ and ReB$_2$ show deep global energy minima at the BABA stacking, yielding high energy barriers for the BAAA and ABBA allotrope (about 0.30-0.45\;eV/at.).

The AAAA$\leftrightarrow$BAAA$\leftrightarrow$BABA$\leftrightarrow$ABBA transformations also lead to volumetric changes, mainly stemming from changes of the $c$ lattice parameter, i.e. relaxations along the [0001] direction.
These can be traced down to the puckering of the boron planes between the metal layers.
Mirroring the dynamical instability of the $\alpha$ structure for the group VI--VII TMB$_2$, boron layers start to pucker already when one of the transition metal layers is slightly shifted from the ideal AAAA stacking.  
WB\textsubscript{2}, TcB\textsubscript{2} and ReB\textsubscript{2} are particularly interesting, since their BAAA, BABA, and ABBA allotropes yield a volume decrease compared to the AAAA stacking, despite of highly puckered boron layers which should, intuitively, take more space than the flat AAAA arrangement.

The relative order of volumes corresponding to the four diboride allotropes also seems to inversely correlate with their bulk moduli, meaning that TMB$_2$ stackings with higher volume exhibit lower bulk modulus. 
The overall highest $B$, $G$, and $E$ moduli are predicted for BABA-ReB$_2$, pointing towards excellent strength.
The calculated Poisson's ratio and the Cauchy pressure indicate changes in ductility when moving from group IV to group VII TMB\textsubscript{2}s but \add{also} depending on the stacking of the transition metal planes.
Specifically, ductility increases with incre\add{a}sing VEC for the $\alpha$-structure and decreases with VEC for the $\gamma$-structure, while a minimum in ductility in the $\omega$-structure is predicted for group V TMB\textsubscript{2}s.
\add{Filling of the $d$ orbitals in higher group TMs weakens the B-B bonds, increases the preference to form strong TM-B bonds, and raises the DOS at $E_F$ in the AAAA allotrope. The combination of previous observations supports structural relaxation that moves TM in line with B to utilizing $d_{z^2}$-$p_z$ hybridization to split the non-bonding $d$\textsubscript{$z$\textsuperscript{$2$}} DOS peak at E\textsubscript{$F$}. The increasing preference for the TM-B interaction over B-B causes boron sheets to pucker to reduce TM-B distance.}

\section*{Acknowledgements}
NK acknowledges the Hertha Firnberg fellowship by Austrian Science Fund, FWF, (T30801).
Computational resources were provided by the Vienna Scientific Cluster (VSC), the cluster at the Montanuniversität Leoben (MUL-hpc), and by Swedish National Infrastructure for Computing (SNIC), on the Clusters located at the National Supercomputer Centre (NSC) in Link\"{o}ping, the Center for High Performance Computing (PDC) in Stockholm, and at the High Performance Computing Center North (HPC2N) in Ume\r{a}, Sweden. \add{This work has been partially supported by the Spanish Ministry of Science and Innovation with PID2019-105488GB-I00 (JJ).}

\input{manuscript_diborides_refs}

\end{document}

%% file: manuscript_diborides_refs.tex
%

%% file: manuscript_diborides.bbl
\begin{thebibliography}{49}%
\makeatletter
\providecommand \@ifxundefined [1]{%
 \@ifx{#1\undefined}
}%
\providecommand \@ifnum [1]{%
 \ifnum #1\expandafter \@firstoftwo
 \else \expandafter \@secondoftwo
 \fi
}%
\providecommand \@ifx [1]{%
 \ifx #1\expandafter \@firstoftwo
 \else \expandafter \@secondoftwo
 \fi
}%
\providecommand \natexlab [1]{#1}%
\providecommand \enquote  [1]{``#1''}%
\providecommand \bibnamefont  [1]{#1}%
\providecommand \bibfnamefont [1]{#1}%
\providecommand \citenamefont [1]{#1}%
\providecommand \href@noop [0]{\@secondoftwo}%
\providecommand \href [0]{\begingroup \@sanitize@url \@href}%
\providecommand \@href[1]{\@@startlink{#1}\@@href}%
\providecommand \@@href[1]{\endgroup#1\@@endlink}%
\providecommand \@sanitize@url [0]{\catcode `\\12\catcode `\$12\catcode
  `\&12\catcode `\#12\catcode `\^12\catcode `\_12\catcode `\%12\relax}%
\providecommand \@@startlink[1]{}%
\providecommand \@@endlink[0]{}%
\providecommand \url  [0]{\begingroup\@sanitize@url \@url }%
\providecommand \@url [1]{\endgroup\@href {#1}{\urlprefix }}%
\providecommand \urlprefix  [0]{URL }%
\providecommand \Eprint [0]{\href }%
\providecommand \doibase [0]{https://doi.org/}%
\providecommand \selectlanguage [0]{\@gobble}%
\providecommand \bibinfo  [0]{\@secondoftwo}%
\providecommand \bibfield  [0]{\@secondoftwo}%
\providecommand \translation [1]{[#1]}%
\providecommand \BibitemOpen [0]{}%
\providecommand \bibitemStop [0]{}%
\providecommand \bibitemNoStop [0]{.\EOS\space}%
\providecommand \EOS [0]{\spacefactor3000\relax}%
\providecommand \BibitemShut  [1]{\csname bibitem#1\endcsname}%
\let\auto@bib@innerbib\@empty
\bibitem [{\citenamefont {Fuger}\ \emph {et~al.}(2019)\citenamefont {Fuger},
  \citenamefont {Moraes}, \citenamefont {Hahn}, \citenamefont {Bolvardi},
  \citenamefont {Polcik}, \citenamefont {Riedl},\ and\ \citenamefont
  {Mayrhofer}}]{fuger2019influence}%
  \BibitemOpen
  \bibfield  {author} {\bibinfo {author} {\bibfnamefont {C.}~\bibnamefont
  {Fuger}}, \bibinfo {author} {\bibfnamefont {V.}~\bibnamefont {Moraes}},
  \bibinfo {author} {\bibfnamefont {R.}~\bibnamefont {Hahn}}, \bibinfo {author}
  {\bibfnamefont {H.}~\bibnamefont {Bolvardi}}, \bibinfo {author}
  {\bibfnamefont {P.}~\bibnamefont {Polcik}}, \bibinfo {author} {\bibfnamefont
  {H.}~\bibnamefont {Riedl}},\ and\ \bibinfo {author} {\bibfnamefont {P.~H.}\
  \bibnamefont {Mayrhofer}},\ }\bibfield  {title} {\bibinfo {title} {Influence
  of {T}antalum on phase stability and mechanical properties of {WB}$_2$},\
  }\href@noop {} {\bibfield  {journal} {\bibinfo  {journal} {MRS
  Communications}\ }\textbf {\bibinfo {volume} {9}},\ \bibinfo {pages} {375}
  (\bibinfo {year} {2019})}\BibitemShut {NoStop}%
\bibitem [{\citenamefont {Fuger}\ \emph {et~al.}(2022)\citenamefont {Fuger},
  \citenamefont {Hahn}, \citenamefont {Zauner}, \citenamefont {Wojcik},
  \citenamefont {Weiss}, \citenamefont {Limbeck}, \citenamefont {Hunold},
  \citenamefont {Polcik},\ and\ \citenamefont {Riedl}}]{fuger2022anisotropic}%
  \BibitemOpen
  \bibfield  {author} {\bibinfo {author} {\bibfnamefont {C.}~\bibnamefont
  {Fuger}}, \bibinfo {author} {\bibfnamefont {R.}~\bibnamefont {Hahn}},
  \bibinfo {author} {\bibfnamefont {L.}~\bibnamefont {Zauner}}, \bibinfo
  {author} {\bibfnamefont {T.}~\bibnamefont {Wojcik}}, \bibinfo {author}
  {\bibfnamefont {M.}~\bibnamefont {Weiss}}, \bibinfo {author} {\bibfnamefont
  {A.}~\bibnamefont {Limbeck}}, \bibinfo {author} {\bibfnamefont
  {O.}~\bibnamefont {Hunold}}, \bibinfo {author} {\bibfnamefont
  {P.}~\bibnamefont {Polcik}},\ and\ \bibinfo {author} {\bibfnamefont
  {H.}~\bibnamefont {Riedl}},\ }\bibfield  {title} {\bibinfo {title}
  {Anisotropic super-hardness of hexagonal {WB}$_{2\pm z}$ thin films},\
  }\href@noop {} {\bibfield  {journal} {\bibinfo  {journal} {Materials Research
  Letters}\ }\textbf {\bibinfo {volume} {10}},\ \bibinfo {pages} {70} (\bibinfo
  {year} {2022})}\BibitemShut {NoStop}%
\bibitem [{\citenamefont {Palisaitis}\ \emph {et~al.}(2022)\citenamefont
  {Palisaitis}, \citenamefont {Dahlqvist}, \citenamefont {Hultman},
  \citenamefont {Petrov}, \citenamefont {Rosen},\ and\ \citenamefont
  {Persson}}]{palisaitis2022nature}%
  \BibitemOpen
  \bibfield  {author} {\bibinfo {author} {\bibfnamefont {J.}~\bibnamefont
  {Palisaitis}}, \bibinfo {author} {\bibfnamefont {M.}~\bibnamefont
  {Dahlqvist}}, \bibinfo {author} {\bibfnamefont {L.}~\bibnamefont {Hultman}},
  \bibinfo {author} {\bibfnamefont {I.}~\bibnamefont {Petrov}}, \bibinfo
  {author} {\bibfnamefont {J.}~\bibnamefont {Rosen}},\ and\ \bibinfo {author}
  {\bibfnamefont {P.~O.}\ \bibnamefont {Persson}},\ }\bibfield  {title}
  {\bibinfo {title} {On the nature of planar defects in transition metal
  diboride line compounds},\ }\href@noop {} {\bibfield  {journal} {\bibinfo
  {journal} {Materialia}\ }\textbf {\bibinfo {volume} {24}},\ \bibinfo {pages}
  {101478} (\bibinfo {year} {2022})}\BibitemShut {NoStop}%
\bibitem [{\citenamefont {Palisaitis}\ \emph {et~al.}(2021)\citenamefont
  {Palisaitis}, \citenamefont {Dahlqvist}, \citenamefont {Hall}, \citenamefont
  {Th{\"o}rnberg}, \citenamefont {Persson}, \citenamefont {Nedfors},
  \citenamefont {Hultman}, \citenamefont {Greene}, \citenamefont {Petrov},
  \citenamefont {Rosen},\ and\ \citenamefont
  {Persson}}]{palisaitis2021unpaired}%
  \BibitemOpen
  \bibfield  {author} {\bibinfo {author} {\bibfnamefont {J.}~\bibnamefont
  {Palisaitis}}, \bibinfo {author} {\bibfnamefont {M.}~\bibnamefont
  {Dahlqvist}}, \bibinfo {author} {\bibfnamefont {A.~J.}\ \bibnamefont {Hall}},
  \bibinfo {author} {\bibfnamefont {J.}~\bibnamefont {Th{\"o}rnberg}}, \bibinfo
  {author} {\bibfnamefont {I.}~\bibnamefont {Persson}}, \bibinfo {author}
  {\bibfnamefont {N.}~\bibnamefont {Nedfors}}, \bibinfo {author} {\bibfnamefont
  {L.}~\bibnamefont {Hultman}}, \bibinfo {author} {\bibfnamefont {J.~E.}\
  \bibnamefont {Greene}}, \bibinfo {author} {\bibfnamefont {I.}~\bibnamefont
  {Petrov}}, \bibinfo {author} {\bibfnamefont {J.}~\bibnamefont {Rosen}},\ and\
  \bibinfo {author} {\bibfnamefont {P.~O.}\ \bibnamefont {Persson}},\
  }\bibfield  {title} {\bibinfo {title} {Where is the unpaired transition metal
  in substoichiometric diboride line compounds?},\ }\href@noop {} {\bibfield
  {journal} {\bibinfo  {journal} {Acta Materialia}\ }\textbf {\bibinfo {volume}
  {204}},\ \bibinfo {pages} {116510} (\bibinfo {year} {2021})}\BibitemShut
  {NoStop}%
\bibitem [{\citenamefont {{\v{S}}roba}\ \emph {et~al.}(2020)\citenamefont
  {{\v{S}}roba}, \citenamefont {Fiantok}, \citenamefont {Truchl{\'y}},
  \citenamefont {Roch}, \citenamefont {Zahoran}, \citenamefont
  {Gran{\v{c}}i{\v{c}}}, \citenamefont {{\v{S}}vec}, \citenamefont {Nagy},
  \citenamefont {Izai}, \citenamefont {K{\'u}{\v{s}}},\ and\ \citenamefont
  {Mikula}}]{vsroba2020structure}%
  \BibitemOpen
  \bibfield  {author} {\bibinfo {author} {\bibfnamefont {V.}~\bibnamefont
  {{\v{S}}roba}}, \bibinfo {author} {\bibfnamefont {T.}~\bibnamefont
  {Fiantok}}, \bibinfo {author} {\bibfnamefont {M.}~\bibnamefont
  {Truchl{\'y}}}, \bibinfo {author} {\bibfnamefont {T.}~\bibnamefont {Roch}},
  \bibinfo {author} {\bibfnamefont {M.}~\bibnamefont {Zahoran}}, \bibinfo
  {author} {\bibfnamefont {B.}~\bibnamefont {Gran{\v{c}}i{\v{c}}}}, \bibinfo
  {author} {\bibfnamefont {P.}~\bibnamefont {{\v{S}}vec}}, \bibinfo {author}
  {\bibfnamefont {{\v{S}}.}~\bibnamefont {Nagy}}, \bibinfo {author}
  {\bibfnamefont {V.}~\bibnamefont {Izai}}, \bibinfo {author} {\bibfnamefont
  {P.}~\bibnamefont {K{\'u}{\v{s}}}},\ and\ \bibinfo {author} {\bibfnamefont
  {M.}~\bibnamefont {Mikula}},\ }\bibfield  {title} {\bibinfo {title}
  {Structure evolution and mechanical properties of hard tantalum diboride
  films},\ }\href@noop {} {\bibfield  {journal} {\bibinfo  {journal} {Journal
  of Vacuum Science \& Technology A: Vacuum, Surfaces, and Films}\ }\textbf
  {\bibinfo {volume} {38}},\ \bibinfo {pages} {033408} (\bibinfo {year}
  {2020})}\BibitemShut {NoStop}%
\bibitem [{\citenamefont {Paul}\ \emph
  {et~al.}(2021{\natexlab{a}})\citenamefont {Paul}, \citenamefont {Okamoto},
  \citenamefont {Kusakari}, \citenamefont {Chen}, \citenamefont {Kishida},
  \citenamefont {Inui},\ and\ \citenamefont {Otani}}]{paul2021plastic}%
  \BibitemOpen
  \bibfield  {author} {\bibinfo {author} {\bibfnamefont {B.}~\bibnamefont
  {Paul}}, \bibinfo {author} {\bibfnamefont {N.~L.}\ \bibnamefont {Okamoto}},
  \bibinfo {author} {\bibfnamefont {M.}~\bibnamefont {Kusakari}}, \bibinfo
  {author} {\bibfnamefont {Z.}~\bibnamefont {Chen}}, \bibinfo {author}
  {\bibfnamefont {K.}~\bibnamefont {Kishida}}, \bibinfo {author} {\bibfnamefont
  {H.}~\bibnamefont {Inui}},\ and\ \bibinfo {author} {\bibfnamefont
  {S.}~\bibnamefont {Otani}},\ }\bibfield  {title} {\bibinfo {title} {Plastic
  deformation of single crystals of {CrB}$_2$, {TiB}$_2$ and {ZrB}$_2$ with the
  hexagonal {AlB}$_2$ structure},\ }\href@noop {} {\bibfield  {journal}
  {\bibinfo  {journal} {Acta Materialia}\ }\textbf {\bibinfo {volume} {211}},\
  \bibinfo {pages} {116857} (\bibinfo {year} {2021}{\natexlab{a}})}\BibitemShut
  {NoStop}%
\bibitem [{\citenamefont {Magnuson}\ \emph {et~al.}(2022)\citenamefont
  {Magnuson}, \citenamefont {Hultman},\ and\ \citenamefont
  {H{\"o}gberg}}]{magnuson2021review}%
  \BibitemOpen
  \bibfield  {author} {\bibinfo {author} {\bibfnamefont {M.}~\bibnamefont
  {Magnuson}}, \bibinfo {author} {\bibfnamefont {L.}~\bibnamefont {Hultman}},\
  and\ \bibinfo {author} {\bibfnamefont {H.}~\bibnamefont {H{\"o}gberg}},\
  }\bibfield  {title} {\bibinfo {title} {Review of transition-metal diboride
  thin films},\ }\href {https://doi.org/10.1016/j.vacuum.2021.110567}
  {\bibfield  {journal} {\bibinfo  {journal} {Vacuum}\ }\textbf {\bibinfo
  {volume} {196}},\ \bibinfo {pages} {110567} (\bibinfo {year}
  {2022})}\BibitemShut {NoStop}%
\bibitem [{\citenamefont {Mitterer}(1997)}]{mitterer1997borides}%
  \BibitemOpen
  \bibfield  {author} {\bibinfo {author} {\bibfnamefont {C.}~\bibnamefont
  {Mitterer}},\ }\bibfield  {title} {\bibinfo {title} {Borides in thin film
  technology},\ }\href@noop {} {\bibfield  {journal} {\bibinfo  {journal}
  {Journal of solid state chemistry}\ }\textbf {\bibinfo {volume} {133}},\
  \bibinfo {pages} {279} (\bibinfo {year} {1997})}\BibitemShut {NoStop}%
\bibitem [{\citenamefont {Liu}\ \emph {et~al.}(2022)\citenamefont {Liu},
  \citenamefont {Gu}, \citenamefont {Zhang}, \citenamefont {Zheng},
  \citenamefont {Ma},\ and\ \citenamefont {Chen}}]{liu2022superhard}%
  \BibitemOpen
  \bibfield  {author} {\bibinfo {author} {\bibfnamefont {C.}~\bibnamefont
  {Liu}}, \bibinfo {author} {\bibfnamefont {X.}~\bibnamefont {Gu}}, \bibinfo
  {author} {\bibfnamefont {K.}~\bibnamefont {Zhang}}, \bibinfo {author}
  {\bibfnamefont {W.}~\bibnamefont {Zheng}}, \bibinfo {author} {\bibfnamefont
  {Y.}~\bibnamefont {Ma}},\ and\ \bibinfo {author} {\bibfnamefont
  {C.}~\bibnamefont {Chen}},\ }\bibfield  {title} {\bibinfo {title} {Superhard
  metallic compound {TaB}$_2$ via crystal orientation resolved strain
  stiffening},\ }\href@noop {} {\bibfield  {journal} {\bibinfo  {journal}
  {Physical Review B}\ }\textbf {\bibinfo {volume} {105}},\ \bibinfo {pages}
  {024105} (\bibinfo {year} {2022})}\BibitemShut {NoStop}%
\bibitem [{\citenamefont {Chung}\ \emph {et~al.}(2007)\citenamefont {Chung},
  \citenamefont {Weinberger}, \citenamefont {Levine}, \citenamefont {Kavner},
  \citenamefont {Yang}, \citenamefont {Tolbert},\ and\ \citenamefont
  {Kaner}}]{ultra_high_hardness_ReB2}%
  \BibitemOpen
  \bibfield  {author} {\bibinfo {author} {\bibfnamefont {H.-Y.}\ \bibnamefont
  {Chung}}, \bibinfo {author} {\bibfnamefont {M.~B.}\ \bibnamefont
  {Weinberger}}, \bibinfo {author} {\bibfnamefont {J.~B.}\ \bibnamefont
  {Levine}}, \bibinfo {author} {\bibfnamefont {A.}~\bibnamefont {Kavner}},
  \bibinfo {author} {\bibfnamefont {J.-M.}\ \bibnamefont {Yang}}, \bibinfo
  {author} {\bibfnamefont {S.~H.}\ \bibnamefont {Tolbert}},\ and\ \bibinfo
  {author} {\bibfnamefont {R.~B.}\ \bibnamefont {Kaner}},\ }\bibfield  {title}
  {\bibinfo {title} {Synthesis of ultra-incompressible superhard rhenium
  diboride at ambient pressure},\ }\href@noop {} {\bibfield  {journal}
  {\bibinfo  {journal} {Science}\ }\textbf {\bibinfo {volume} {316}},\ \bibinfo
  {pages} {436} (\bibinfo {year} {2007})}\BibitemShut {NoStop}%
\bibitem [{\citenamefont {Nagamatsu}\ \emph {et~al.}(2001)\citenamefont
  {Nagamatsu}, \citenamefont {Nakagawa}, \citenamefont {Muranaka},
  \citenamefont {Zenitani},\ and\ \citenamefont
  {Akimitsu}}]{nagamatsu2001superconductivity}%
  \BibitemOpen
  \bibfield  {author} {\bibinfo {author} {\bibfnamefont {J.}~\bibnamefont
  {Nagamatsu}}, \bibinfo {author} {\bibfnamefont {N.}~\bibnamefont {Nakagawa}},
  \bibinfo {author} {\bibfnamefont {T.}~\bibnamefont {Muranaka}}, \bibinfo
  {author} {\bibfnamefont {Y.}~\bibnamefont {Zenitani}},\ and\ \bibinfo
  {author} {\bibfnamefont {J.}~\bibnamefont {Akimitsu}},\ }\bibfield  {title}
  {\bibinfo {title} {Superconductivity at {39\;K} in magnesium diboride},\
  }\href@noop {} {\bibfield  {journal} {\bibinfo  {journal} {Nature}\ }\textbf
  {\bibinfo {volume} {410}},\ \bibinfo {pages} {63} (\bibinfo {year}
  {2001})}\BibitemShut {NoStop}%
\bibitem [{\citenamefont {Hellgren}\ \emph {et~al.}(2022)\citenamefont
  {Hellgren}, \citenamefont {Sredenschek}, \citenamefont {Petruins},
  \citenamefont {Palisaitis}, \citenamefont {Klimashin}, \citenamefont
  {Sortica}, \citenamefont {Hultman}, \citenamefont {Persson},\ and\
  \citenamefont {Rosen}}]{hellgren2022synthesis}%
  \BibitemOpen
  \bibfield  {author} {\bibinfo {author} {\bibfnamefont {N.}~\bibnamefont
  {Hellgren}}, \bibinfo {author} {\bibfnamefont {A.}~\bibnamefont
  {Sredenschek}}, \bibinfo {author} {\bibfnamefont {A.}~\bibnamefont
  {Petruins}}, \bibinfo {author} {\bibfnamefont {J.}~\bibnamefont
  {Palisaitis}}, \bibinfo {author} {\bibfnamefont {F.~F.}\ \bibnamefont
  {Klimashin}}, \bibinfo {author} {\bibfnamefont {M.~A.}\ \bibnamefont
  {Sortica}}, \bibinfo {author} {\bibfnamefont {L.}~\bibnamefont {Hultman}},
  \bibinfo {author} {\bibfnamefont {P.~O.}\ \bibnamefont {Persson}},\ and\
  \bibinfo {author} {\bibfnamefont {J.}~\bibnamefont {Rosen}},\ }\bibfield
  {title} {\bibinfo {title} {Synthesis and characterization of {TiB}$_x$
  (1.2$\leq x\leq$2.8) thin films grown by {DC} magnetron co-sputtering from
  {TiB}$_2$ and {Ti} targets},\ }\href@noop {} {\bibfield  {journal} {\bibinfo
  {journal} {Surface and Coatings Technology}\ }\textbf {\bibinfo {volume}
  {433}},\ \bibinfo {pages} {128110} (\bibinfo {year} {2022})}\BibitemShut
  {NoStop}%
\bibitem [{\citenamefont {Bakhit}\ \emph {et~al.}(2020)\citenamefont {Bakhit},
  \citenamefont {Palisaitis}, \citenamefont {Th{\"o}rnberg}, \citenamefont
  {Rosen}, \citenamefont {Persson}, \citenamefont {Hultman}, \citenamefont
  {Petrov}, \citenamefont {Greene},\ and\ \citenamefont
  {Greczynski}}]{bakhit2020improving}%
  \BibitemOpen
  \bibfield  {author} {\bibinfo {author} {\bibfnamefont {B.}~\bibnamefont
  {Bakhit}}, \bibinfo {author} {\bibfnamefont {J.}~\bibnamefont {Palisaitis}},
  \bibinfo {author} {\bibfnamefont {J.}~\bibnamefont {Th{\"o}rnberg}}, \bibinfo
  {author} {\bibfnamefont {J.}~\bibnamefont {Rosen}}, \bibinfo {author}
  {\bibfnamefont {P.~O.}\ \bibnamefont {Persson}}, \bibinfo {author}
  {\bibfnamefont {L.}~\bibnamefont {Hultman}}, \bibinfo {author} {\bibfnamefont
  {I.}~\bibnamefont {Petrov}}, \bibinfo {author} {\bibfnamefont {J.~E.}\
  \bibnamefont {Greene}},\ and\ \bibinfo {author} {\bibfnamefont
  {G.}~\bibnamefont {Greczynski}},\ }\bibfield  {title} {\bibinfo {title}
  {Improving the high-temperature oxidation resistance of {TiB}$_2$ thin films
  by alloying with al},\ }\href@noop {} {\bibfield  {journal} {\bibinfo
  {journal} {Acta Materialia}\ }\textbf {\bibinfo {volume} {196}},\ \bibinfo
  {pages} {677} (\bibinfo {year} {2020})}\BibitemShut {NoStop}%
\bibitem [{\citenamefont {Munro}(2000)}]{munro2000material}%
  \BibitemOpen
  \bibfield  {author} {\bibinfo {author} {\bibfnamefont {R.~G.}\ \bibnamefont
  {Munro}},\ }\bibfield  {title} {\bibinfo {title} {Material properties of
  titanium diboride},\ }\href@noop {} {\bibfield  {journal} {\bibinfo
  {journal} {Journal of Research of the National institute of standards and
  Technology}\ }\textbf {\bibinfo {volume} {105}},\ \bibinfo {pages} {709}
  (\bibinfo {year} {2000})}\BibitemShut {NoStop}%
\bibitem [{\citenamefont {Wang~C}(2020)}]{properties_WB2}%
  \BibitemOpen
  \bibfield  {author} {\bibinfo {author} {\bibfnamefont {X.~Y.}\ \bibnamefont
  {Wang~C}, \bibfnamefont {Song~L}},\ }\bibfield  {title} {\bibinfo {title}
  {Mechanical and electrical characteristics of wb\textsubscript{2} synthesized
  at high pressure and high temperature},\ }\href@noop {} {\bibfield  {journal}
  {\bibinfo  {journal} {Materials (Basel)}\ }\textbf {\bibinfo {volume} {13}},\
  \bibinfo {pages} {1212} (\bibinfo {year} {2020})}\BibitemShut {NoStop}%
\bibitem [{\citenamefont {La~Placa}\ and\ \citenamefont
  {Post}(1962)}]{La_Placa1962-pj}%
  \BibitemOpen
  \bibfield  {author} {\bibinfo {author} {\bibfnamefont {S.~J.}\ \bibnamefont
  {La~Placa}}\ and\ \bibinfo {author} {\bibfnamefont {B.}~\bibnamefont
  {Post}},\ }\bibfield  {title} {\bibinfo {title} {The crystal structure of
  rhenium diboride},\ }\href {https://doi.org/10.1107/S0365110X62000298}
  {\bibfield  {journal} {\bibinfo  {journal} {Acta Crystallogr.}\ }\textbf
  {\bibinfo {volume} {15}},\ \bibinfo {pages} {97} (\bibinfo {year}
  {1962})}\BibitemShut {NoStop}%
\bibitem [{\citenamefont {Moraes}\ \emph {et~al.}(2018)\citenamefont {Moraes},
  \citenamefont {Riedl}, \citenamefont {Fuger}, \citenamefont {Polcik},
  \citenamefont {Bolvardi}, \citenamefont {Holec},\ and\ \citenamefont
  {Mayrhofer}}]{moraes2018ab}%
  \BibitemOpen
  \bibfield  {author} {\bibinfo {author} {\bibfnamefont {V.}~\bibnamefont
  {Moraes}}, \bibinfo {author} {\bibfnamefont {H.}~\bibnamefont {Riedl}},
  \bibinfo {author} {\bibfnamefont {C.}~\bibnamefont {Fuger}}, \bibinfo
  {author} {\bibfnamefont {P.}~\bibnamefont {Polcik}}, \bibinfo {author}
  {\bibfnamefont {H.}~\bibnamefont {Bolvardi}}, \bibinfo {author}
  {\bibfnamefont {D.}~\bibnamefont {Holec}},\ and\ \bibinfo {author}
  {\bibfnamefont {P.~H.}\ \bibnamefont {Mayrhofer}},\ }\bibfield  {title}
  {\bibinfo {title} {Ab initio inspired design of ternary boride thin films},\
  }\href@noop {} {\bibfield  {journal} {\bibinfo  {journal} {Scientific
  Reports}\ }\textbf {\bibinfo {volume} {8}},\ \bibinfo {pages} {1} (\bibinfo
  {year} {2018})}\BibitemShut {NoStop}%
\bibitem [{\citenamefont {Kiessling}\ \emph {et~al.}(1947)\citenamefont
  {Kiessling}, \citenamefont {Wetterholm}, \citenamefont {Sill{\'e}n},
  \citenamefont {Linnasalmi},\ and\ \citenamefont
  {Laukkanen}}]{Kiessling1947-sb}%
  \BibitemOpen
  \bibfield  {author} {\bibinfo {author} {\bibfnamefont {R.}~\bibnamefont
  {Kiessling}}, \bibinfo {author} {\bibfnamefont {A.}~\bibnamefont
  {Wetterholm}}, \bibinfo {author} {\bibfnamefont {L.~G.}\ \bibnamefont
  {Sill{\'e}n}}, \bibinfo {author} {\bibfnamefont {A.}~\bibnamefont
  {Linnasalmi}},\ and\ \bibinfo {author} {\bibfnamefont {P.}~\bibnamefont
  {Laukkanen}},\ }\bibfield  {title} {\bibinfo {title} {The crystal structures
  of molybdenum and tungsten borides},\ }\href
  {https://doi.org/10.3891/acta.chem.scand.01-0893} {\bibfield  {journal}
  {\bibinfo  {journal} {Acta Chem. Scand.}\ }\textbf {\bibinfo {volume} {1}},\
  \bibinfo {pages} {893} (\bibinfo {year} {1947})}\BibitemShut {NoStop}%
\bibitem [{\citenamefont {Zhang}\ \emph {et~al.}(2010)\citenamefont {Zhang},
  \citenamefont {Legut}, \citenamefont {Niewa}, \citenamefont {Argon},\ and\
  \citenamefont {Veprek}}]{Zhang2010-ah}%
  \BibitemOpen
  \bibfield  {author} {\bibinfo {author} {\bibfnamefont {R.~F.}\ \bibnamefont
  {Zhang}}, \bibinfo {author} {\bibfnamefont {D.}~\bibnamefont {Legut}},
  \bibinfo {author} {\bibfnamefont {R.}~\bibnamefont {Niewa}}, \bibinfo
  {author} {\bibfnamefont {A.~S.}\ \bibnamefont {Argon}},\ and\ \bibinfo
  {author} {\bibfnamefont {S.}~\bibnamefont {Veprek}},\ }\bibfield  {title}
  {\bibinfo {title} {Shear-induced structural transformation and plasticity in
  ultraincompressible {ReB$_2$} limit its hardness},\ }\href
  {https://doi.org/10.1103/PhysRevB.82.104104} {\bibfield  {journal} {\bibinfo
  {journal} {Phys. Rev. B Condens. Matter}\ }\textbf {\bibinfo {volume} {82}},\
  \bibinfo {pages} {104104} (\bibinfo {year} {2010})}\BibitemShut {NoStop}%
\bibitem [{\citenamefont {Hunter}\ \emph {et~al.}(2016)\citenamefont {Hunter},
  \citenamefont {Yu}, \citenamefont {De~Leon}, \citenamefont {Weinberger},
  \citenamefont {Fahrenholtz}, \citenamefont {Hilmas}, \citenamefont {Weaver},\
  and\ \citenamefont {Thompson}}]{Hunter2016-ee}%
  \BibitemOpen
  \bibfield  {author} {\bibinfo {author} {\bibfnamefont {B.}~\bibnamefont
  {Hunter}}, \bibinfo {author} {\bibfnamefont {X.-X.}\ \bibnamefont {Yu}},
  \bibinfo {author} {\bibfnamefont {N.}~\bibnamefont {De~Leon}}, \bibinfo
  {author} {\bibfnamefont {C.}~\bibnamefont {Weinberger}}, \bibinfo {author}
  {\bibfnamefont {W.}~\bibnamefont {Fahrenholtz}}, \bibinfo {author}
  {\bibfnamefont {G.}~\bibnamefont {Hilmas}}, \bibinfo {author} {\bibfnamefont
  {M.~L.}\ \bibnamefont {Weaver}},\ and\ \bibinfo {author} {\bibfnamefont
  {G.~B.}\ \bibnamefont {Thompson}},\ }\bibfield  {title} {\bibinfo {title}
  {Investigations into the slip behavior of zirconium diboride},\ }\href
  {https://doi.org/10.1557/jmr.2016.201} {\bibfield  {journal} {\bibinfo
  {journal} {J. Mater. Res.}\ }\textbf {\bibinfo {volume} {31}},\ \bibinfo
  {pages} {2749} (\bibinfo {year} {2016})}\BibitemShut {NoStop}%
\bibitem [{\citenamefont {Paul}\ \emph
  {et~al.}(2021{\natexlab{b}})\citenamefont {Paul}, \citenamefont {Okamoto},
  \citenamefont {Kusakari}, \citenamefont {Chen}, \citenamefont {Kishida},
  \citenamefont {Inui},\ and\ \citenamefont {Otani}}]{Paul2021-pd}%
  \BibitemOpen
  \bibfield  {author} {\bibinfo {author} {\bibfnamefont {B.}~\bibnamefont
  {Paul}}, \bibinfo {author} {\bibfnamefont {N.~L.}\ \bibnamefont {Okamoto}},
  \bibinfo {author} {\bibfnamefont {M.}~\bibnamefont {Kusakari}}, \bibinfo
  {author} {\bibfnamefont {Z.}~\bibnamefont {Chen}}, \bibinfo {author}
  {\bibfnamefont {K.}~\bibnamefont {Kishida}}, \bibinfo {author} {\bibfnamefont
  {H.}~\bibnamefont {Inui}},\ and\ \bibinfo {author} {\bibfnamefont
  {S.}~\bibnamefont {Otani}},\ }\bibfield  {title} {\bibinfo {title} {Plastic
  deformation of single crystals of {CrB$_2$}, {TiB$_2$} and {ZrB$_2$} with the
  hexagonal {AlB$_2$} structure},\ }\href
  {https://doi.org/10.1016/j.actamat.2021.116857} {\bibfield  {journal}
  {\bibinfo  {journal} {Acta Mater.}\ }\textbf {\bibinfo {volume} {211}},\
  \bibinfo {pages} {116857} (\bibinfo {year} {2021}{\natexlab{b}})}\BibitemShut
  {NoStop}%
\bibitem [{\citenamefont {Kresse}\ and\ \citenamefont
  {Furthm\"uller}(1996)}]{Kresse1996Efficient}%
  \BibitemOpen
  \bibfield  {author} {\bibinfo {author} {\bibfnamefont {G.}~\bibnamefont
  {Kresse}}\ and\ \bibinfo {author} {\bibfnamefont {J.}~\bibnamefont
  {Furthm\"uller}},\ }\bibfield  {title} {\bibinfo {title} {Efficient iterative
  schemes for \textit{ab initio} total-energy calculations using a plane-wave
  basis set},\ }\href@noop {} {\bibfield  {journal} {\bibinfo  {journal}
  {Physical Review B}\ }\textbf {\bibinfo {volume} {54}},\ \bibinfo {pages}
  {11169} (\bibinfo {year} {1996})}\BibitemShut {NoStop}%
\bibitem [{\citenamefont {Kresse}\ and\ \citenamefont
  {Joubert}(1999)}]{Kresse1999From}%
  \BibitemOpen
  \bibfield  {author} {\bibinfo {author} {\bibfnamefont {G.}~\bibnamefont
  {Kresse}}\ and\ \bibinfo {author} {\bibfnamefont {D.}~\bibnamefont
  {Joubert}},\ }\bibfield  {title} {\bibinfo {title} {From ultrasoft
  pseudopotentials to the projector augmented-wave method},\ }\href@noop {}
  {\bibfield  {journal} {\bibinfo  {journal} {Physical Review B}\ }\textbf
  {\bibinfo {volume} {59}},\ \bibinfo {pages} {1758} (\bibinfo {year}
  {1999})}\BibitemShut {NoStop}%
\bibitem [{\citenamefont {Perdew}\ \emph {et~al.}(1996)\citenamefont {Perdew},
  \citenamefont {Burke},\ and\ \citenamefont
  {Ernzerhof}}]{perdew1996generalized}%
  \BibitemOpen
  \bibfield  {author} {\bibinfo {author} {\bibfnamefont {J.~P.}\ \bibnamefont
  {Perdew}}, \bibinfo {author} {\bibfnamefont {K.}~\bibnamefont {Burke}},\ and\
  \bibinfo {author} {\bibfnamefont {M.}~\bibnamefont {Ernzerhof}},\ }\bibfield
  {title} {\bibinfo {title} {Generalized gradient approximation made simple},\
  }\href@noop {} {\bibfield  {journal} {\bibinfo  {journal} {Physical review
  letters}\ }\textbf {\bibinfo {volume} {77}},\ \bibinfo {pages} {3865}
  (\bibinfo {year} {1996})}\BibitemShut {NoStop}%
\bibitem [{\citenamefont {Bu{\v{c}}ko}\ \emph {et~al.}(2005)\citenamefont
  {Bu{\v{c}}ko}, \citenamefont {Hafner},\ and\ \citenamefont
  {{\'A}ngy{\'a}n}}]{buvcko2005geometry}%
  \BibitemOpen
  \bibfield  {author} {\bibinfo {author} {\bibfnamefont {T.}~\bibnamefont
  {Bu{\v{c}}ko}}, \bibinfo {author} {\bibfnamefont {J.}~\bibnamefont
  {Hafner}},\ and\ \bibinfo {author} {\bibfnamefont {J.}~\bibnamefont
  {{\'A}ngy{\'a}n}},\ }\bibfield  {title} {\bibinfo {title} {Geometry
  optimization of periodic systems using internal coordinates},\ }\href@noop {}
  {\bibfield  {journal} {\bibinfo  {journal} {The Journal of Chemical Physics}\
  }\textbf {\bibinfo {volume} {122}},\ \bibinfo {pages} {124508} (\bibinfo
  {year} {2005})}\BibitemShut {NoStop}%
\bibitem [{\citenamefont {Le~Page}\ and\ \citenamefont
  {Saxe}(2002)}]{le2002symmetry}%
  \BibitemOpen
  \bibfield  {author} {\bibinfo {author} {\bibfnamefont {Y.}~\bibnamefont
  {Le~Page}}\ and\ \bibinfo {author} {\bibfnamefont {P.}~\bibnamefont {Saxe}},\
  }\bibfield  {title} {\bibinfo {title} {Symmetry-general least-squares
  extraction of elastic data for strained materials from ab initio calculations
  of stress},\ }\href@noop {} {\bibfield  {journal} {\bibinfo  {journal}
  {Physical Review B}\ }\textbf {\bibinfo {volume} {65}},\ \bibinfo {pages}
  {104104} (\bibinfo {year} {2002})}\BibitemShut {NoStop}%
\bibitem [{\citenamefont {Le~P.}\ and\ \citenamefont
  {Saxe}(2001)}]{le2001symmetry}%
  \BibitemOpen
  \bibfield  {author} {\bibinfo {author} {\bibfnamefont {Y.}~\bibnamefont
  {Le~P.}}\ and\ \bibinfo {author} {\bibfnamefont {P.}~\bibnamefont {Saxe}},\
  }\bibfield  {title} {\bibinfo {title} {Symmetry-general least-squares
  extraction of elastic coefficients from {\it{ab initio}} total energy
  calculations},\ }\href@noop {} {\bibfield  {journal} {\bibinfo  {journal}
  {Physical Review B}\ }\textbf {\bibinfo {volume} {63}},\ \bibinfo {pages}
  {174103} (\bibinfo {year} {2001})}\BibitemShut {NoStop}%
\bibitem [{\citenamefont {Yu}\ \emph {et~al.}(2010)\citenamefont {Yu},
  \citenamefont {Zhu},\ and\ \citenamefont {Ye}}]{Yu2010-vr}%
  \BibitemOpen
  \bibfield  {author} {\bibinfo {author} {\bibfnamefont {R.}~\bibnamefont
  {Yu}}, \bibinfo {author} {\bibfnamefont {J.}~\bibnamefont {Zhu}},\ and\
  \bibinfo {author} {\bibfnamefont {H.~Q.}\ \bibnamefont {Ye}},\ }\bibfield
  {title} {\bibinfo {title} {Calculations of single-crystal elastic constants
  made simple},\ }\href {https://doi.org/10.1016/j.cpc.2009.11.017} {\bibfield
  {journal} {\bibinfo  {journal} {Comput. Phys. Commun.}\ }\textbf {\bibinfo
  {volume} {181}},\ \bibinfo {pages} {671} (\bibinfo {year}
  {2010})}\BibitemShut {NoStop}%
\bibitem [{\citenamefont {Mouhat}\ and\ \citenamefont
  {Coudert}(2014)}]{mouhat2014necessary}%
  \BibitemOpen
  \bibfield  {author} {\bibinfo {author} {\bibfnamefont {F.}~\bibnamefont
  {Mouhat}}\ and\ \bibinfo {author} {\bibfnamefont {F.-X.}\ \bibnamefont
  {Coudert}},\ }\bibfield  {title} {\bibinfo {title} {Necessary and sufficient
  elastic stability conditions in various crystal systems},\ }\href@noop {}
  {\bibfield  {journal} {\bibinfo  {journal} {Physical Review B}\ }\textbf
  {\bibinfo {volume} {90}},\ \bibinfo {pages} {224104} (\bibinfo {year}
  {2014})}\BibitemShut {NoStop}%
\bibitem [{\citenamefont {Nye}(1985)}]{nye1985physical}%
  \BibitemOpen
  \bibfield  {author} {\bibinfo {author} {\bibfnamefont {J.~F.}\ \bibnamefont
  {Nye}},\ }\href@noop {} {\emph {\bibinfo {title} {Physical properties of
  crystals: their representation by tensors and matrices}}}\ (\bibinfo
  {publisher} {Oxford University Press},\ \bibinfo {year} {1985})\BibitemShut
  {NoStop}%
\bibitem [{\citenamefont {Hill}(1952)}]{hill1952elastic}%
  \BibitemOpen
  \bibfield  {author} {\bibinfo {author} {\bibfnamefont {R.}~\bibnamefont
  {Hill}},\ }\bibfield  {title} {\bibinfo {title} {The elastic behaviour of a
  crystalline aggregate},\ }\href@noop {} {\bibfield  {journal} {\bibinfo
  {journal} {Proceedings of the Physical Society. Section A}\ }\textbf
  {\bibinfo {volume} {65}},\ \bibinfo {pages} {349} (\bibinfo {year}
  {1952})}\BibitemShut {NoStop}%
\bibitem [{\citenamefont {Togo}\ \emph {et~al.}(2008)\citenamefont {Togo},
  \citenamefont {Oba},\ and\ \citenamefont {Tanaka}}]{togo2008first}%
  \BibitemOpen
  \bibfield  {author} {\bibinfo {author} {\bibfnamefont {A.}~\bibnamefont
  {Togo}}, \bibinfo {author} {\bibfnamefont {F.}~\bibnamefont {Oba}},\ and\
  \bibinfo {author} {\bibfnamefont {I.}~\bibnamefont {Tanaka}},\ }\bibfield
  {title} {\bibinfo {title} {First-principles calculations of the ferroelastic
  transition between rutile-type and {CaCl$_2$}-type {SiO$_2$} at high
  pressures},\ }\href@noop {} {\bibfield  {journal} {\bibinfo  {journal}
  {Physical Review B}\ }\textbf {\bibinfo {volume} {78}},\ \bibinfo {pages}
  {134106} (\bibinfo {year} {2008})}\BibitemShut {NoStop}%
\bibitem [{\citenamefont {Dronskowski}\ and\ \citenamefont
  {Bl{\"o}chl}(1993)}]{dronskowski1993crystal}%
  \BibitemOpen
  \add{
  \bibfield  {author} {\bibinfo {author} {\bibfnamefont {R.}~\bibnamefont
  {Dronskowski}}\ and\ \bibinfo {author} {\bibfnamefont {P.~E.}\ \bibnamefont
  {Bl{\"o}chl}},\ }\bibfield  {title} {\bibinfo {title} {Crystal orbital
  hamilton populations ({COHP}): energy-resolved visualization of chemical
  bonding in solids based on density-functional calculations},\ }\href@noop {}
  {\bibfield  {journal} {\bibinfo  {journal} {The Journal of Physical
  Chemistry}\ }\textbf {\bibinfo {volume} {97}},\ \bibinfo {pages} {8617}
  (\bibinfo {year} {1993})}\BibitemShut {NoStop}%
  }
\bibitem [{\citenamefont {Maintz}\ \emph {et~al.}(2013)\citenamefont {Maintz},
  \citenamefont {Deringer}, \citenamefont {Tchougr{\'e}eff},\ and\
  \citenamefont {Dronskowski}}]{maintz2013analytic}%
  \BibitemOpen
  \add{
  \bibfield  {author} {\bibinfo {author} {\bibfnamefont {S.}~\bibnamefont
  {Maintz}}, \bibinfo {author} {\bibfnamefont {V.~L.}\ \bibnamefont
  {Deringer}}, \bibinfo {author} {\bibfnamefont {A.~L.}\ \bibnamefont
  {Tchougr{\'e}eff}},\ and\ \bibinfo {author} {\bibfnamefont {R.}~\bibnamefont
  {Dronskowski}},\ }\bibfield  {title} {\bibinfo {title} {Analytic projection
  from plane-wave and {PAW} wavefunctions and application to chemical-bonding
  analysis in solids},\ }\href@noop {} {\bibfield  {journal} {\bibinfo
  {journal} {Journal of Computational Chemistry}\ }\textbf {\bibinfo {volume}
  {34}},\ \bibinfo {pages} {2557} (\bibinfo {year} {2013})}\BibitemShut
  {NoStop}%
  }
\bibitem [{\citenamefont {Maintz}\ \emph {et~al.}(2016)\citenamefont {Maintz},
  \citenamefont {Deringer}, \citenamefont {Tchougr{\'e}eff},\ and\
  \citenamefont {Dronskowski}}]{maintz2016lobster}%
  \BibitemOpen
  \add{
  \bibfield  {author} {\bibinfo {author} {\bibfnamefont {S.}~\bibnamefont
  {Maintz}}, \bibinfo {author} {\bibfnamefont {V.~L.}\ \bibnamefont
  {Deringer}}, \bibinfo {author} {\bibfnamefont {A.~L.}\ \bibnamefont
  {Tchougr{\'e}eff}},\ and\ \bibinfo {author} {\bibfnamefont {R.}~\bibnamefont
  {Dronskowski}},\ }\href@noop {} {\bibinfo {title} {{LOBSTER}: A tool to
  extract chemical bonding from plane-wave based {DFT}}} (\bibinfo {year}
  {2016})\BibitemShut {NoStop}%
  }
\bibitem [{\citenamefont {Ding}\ \emph {et~al.}(2016)\citenamefont {Ding},
  \citenamefont {Shao}, \citenamefont {Zhang}, \citenamefont {Lu},
  \citenamefont {Ding}, \citenamefont {Ning},\ and\ \citenamefont
  {Huang}}]{MoB2_WB2_structures}%
  \BibitemOpen
  \bibfield  {author} {\bibinfo {author} {\bibfnamefont {L.-P.}\ \bibnamefont
  {Ding}}, \bibinfo {author} {\bibfnamefont {P.}~\bibnamefont {Shao}}, \bibinfo
  {author} {\bibfnamefont {F.-H.}\ \bibnamefont {Zhang}}, \bibinfo {author}
  {\bibfnamefont {C.}~\bibnamefont {Lu}}, \bibinfo {author} {\bibfnamefont
  {L.}~\bibnamefont {Ding}}, \bibinfo {author} {\bibfnamefont {S.~Y.}\
  \bibnamefont {Ning}},\ and\ \bibinfo {author} {\bibfnamefont {X.~F.}\
  \bibnamefont {Huang}},\ }\bibfield  {title} {\bibinfo {title} {Crystal
  structures, stabilities, electronic properties, and hardness of {MoB$_2$}:
  First-principles calculations},\ }\href
  {https://doi.org/10.1021/acs.inorgchem.6b00899} {\bibfield  {journal}
  {\bibinfo  {journal} {Inorganic Chemistry}\ }\textbf {\bibinfo {volume}
  {55}},\ \bibinfo {pages} {7033} (\bibinfo {year} {2016})}\BibitemShut
  {NoStop}%
\bibitem [{\citenamefont {The
  Materials~Project}(2020{\natexlab{a}})}]{osti_1207357}%
  \BibitemOpen
  \add{
  \bibfield  {author} {\bibinfo {author} {\bibnamefont {The
  Materials~Project}},\ }\bibfield  {title} {\bibinfo {title} {Materials data
  on {CrB$_2$} by {Materials Project}}\ }\href
  {https://doi.org/10.17188/1207357} {10.17188/1207357} (\bibinfo {year}
  {2020}{\natexlab{a}})\BibitemShut {NoStop}%
  }
\bibitem [{\citenamefont {The
  Materials~Project}(2020{\natexlab{b}})}]{osti_1277089}%
  \BibitemOpen
  \add{
  \bibfield  {author} {\bibinfo {author} {\bibnamefont {The
  Materials~Project}},\ }\bibfield  {title} {\bibinfo {title} {Materials data
  on {MnB$_2$} by {Materials Project}}\ }\href
  {https://doi.org/10.17188/1277089} {10.17188/1277089} (\bibinfo {year}
  {2020}{\natexlab{b}})\BibitemShut {NoStop}%
  }
\bibitem [{\citenamefont {Hao}\ \emph {et~al.}(2006)\citenamefont {Hao},
  \citenamefont {Xu}, \citenamefont {Wu}, \citenamefont {Zhou}, \citenamefont
  {Liu}, \citenamefont {Cao},\ and\ \citenamefont {Meng}}]{ReB2_WB2_paper}%
  \BibitemOpen
  \bibfield  {author} {\bibinfo {author} {\bibfnamefont {X.}~\bibnamefont
  {Hao}}, \bibinfo {author} {\bibfnamefont {Y.}~\bibnamefont {Xu}}, \bibinfo
  {author} {\bibfnamefont {Z.}~\bibnamefont {Wu}}, \bibinfo {author}
  {\bibfnamefont {D.}~\bibnamefont {Zhou}}, \bibinfo {author} {\bibfnamefont
  {X.}~\bibnamefont {Liu}}, \bibinfo {author} {\bibfnamefont {X.}~\bibnamefont
  {Cao}},\ and\ \bibinfo {author} {\bibfnamefont {J.}~\bibnamefont {Meng}},\
  }\bibfield  {title} {\bibinfo {title} {Low-compressibility and hard materials
  $\mathrm{Re}{\mathrm{b}}_{2}$ and $\mathrm{W}{\mathrm{b}}_{2}$: Prediction
  from first-principles study},\ }\href
  {https://doi.org/10.1103/PhysRevB.74.224112} {\bibfield  {journal} {\bibinfo
  {journal} {Phys. Rev. B}\ }\textbf {\bibinfo {volume} {74}},\ \bibinfo
  {pages} {224112} (\bibinfo {year} {2006})}\BibitemShut {NoStop}%
\bibitem [{\citenamefont {Levine}\ \emph {et~al.}(2010)\citenamefont {Levine},
  \citenamefont {Betts}, \citenamefont {Garrett}, \citenamefont {Guo},
  \citenamefont {Eng}, \citenamefont {Migliori},\ and\ \citenamefont
  {Kaner}}]{levine2010full}%
  \BibitemOpen
  \bibfield  {author} {\bibinfo {author} {\bibfnamefont {J.}~\bibnamefont
  {Levine}}, \bibinfo {author} {\bibfnamefont {J.}~\bibnamefont {Betts}},
  \bibinfo {author} {\bibfnamefont {J.}~\bibnamefont {Garrett}}, \bibinfo
  {author} {\bibfnamefont {S.}~\bibnamefont {Guo}}, \bibinfo {author}
  {\bibfnamefont {J.}~\bibnamefont {Eng}}, \bibinfo {author} {\bibfnamefont
  {A.}~\bibnamefont {Migliori}},\ and\ \bibinfo {author} {\bibfnamefont
  {R.}~\bibnamefont {Kaner}},\ }\bibfield  {title} {\bibinfo {title} {Full
  elastic tensor of a crystal of the superhard compound {ReB$_2$}},\
  }\href@noop {} {\bibfield  {journal} {\bibinfo  {journal} {Acta Materialia}\
  }\textbf {\bibinfo {volume} {58}},\ \bibinfo {pages} {1530} (\bibinfo {year}
  {2010})}\BibitemShut {NoStop}%
\bibitem [{\citenamefont {Gu}\ \emph {et~al.}(2021)\citenamefont {Gu},
  \citenamefont {Liu}, \citenamefont {Guo}, \citenamefont {Zhang},\ and\
  \citenamefont {Chen}}]{gu2021sorting}%
  \BibitemOpen
  \bibfield  {author} {\bibinfo {author} {\bibfnamefont {X.}~\bibnamefont
  {Gu}}, \bibinfo {author} {\bibfnamefont {C.}~\bibnamefont {Liu}}, \bibinfo
  {author} {\bibfnamefont {H.}~\bibnamefont {Guo}}, \bibinfo {author}
  {\bibfnamefont {K.}~\bibnamefont {Zhang}},\ and\ \bibinfo {author}
  {\bibfnamefont {C.}~\bibnamefont {Chen}},\ }\bibfield  {title} {\bibinfo
  {title} {Sorting transition-metal diborides: New descriptor for mechanical
  properties},\ }\href@noop {} {\bibfield  {journal} {\bibinfo  {journal} {Acta
  Materialia}\ }\textbf {\bibinfo {volume} {207}},\ \bibinfo {pages} {116685}
  (\bibinfo {year} {2021})}\BibitemShut {NoStop}%
\bibitem [{\citenamefont {Greaves}\ \emph {et~al.}(2011)\citenamefont
  {Greaves}, \citenamefont {Greer}, \citenamefont {Lakes},\ and\ \citenamefont
  {Rouxel}}]{greaves2011poisson}%
  \BibitemOpen
  \bibfield  {author} {\bibinfo {author} {\bibfnamefont {G.~N.}\ \bibnamefont
  {Greaves}}, \bibinfo {author} {\bibfnamefont {A.~L.}\ \bibnamefont {Greer}},
  \bibinfo {author} {\bibfnamefont {R.~S.}\ \bibnamefont {Lakes}},\ and\
  \bibinfo {author} {\bibfnamefont {T.}~\bibnamefont {Rouxel}},\ }\bibfield
  {title} {\bibinfo {title} {Poisson's ratio and modern materials},\
  }\href@noop {} {\bibfield  {journal} {\bibinfo  {journal} {Nature Materials}\
  }\textbf {\bibinfo {volume} {10}},\ \bibinfo {pages} {823} (\bibinfo {year}
  {2011})}\BibitemShut {NoStop}%
\bibitem [{\citenamefont {Pettifor}(1992)}]{pettifor1992theoretical}%
  \BibitemOpen
  \bibfield  {author} {\bibinfo {author} {\bibfnamefont {D.}~\bibnamefont
  {Pettifor}},\ }\bibfield  {title} {\bibinfo {title} {Theoretical predictions
  of structure and related properties of intermetallics},\ }\href@noop {}
  {\bibfield  {journal} {\bibinfo  {journal} {Materials Science and
  Technology}\ }\textbf {\bibinfo {volume} {8}},\ \bibinfo {pages} {345}
  (\bibinfo {year} {1992})}\BibitemShut {NoStop}%
\bibitem [{\citenamefont {Koutn{\'a}}\ \emph {et~al.}(2021)\citenamefont
  {Koutn{\'a}}, \citenamefont {Brenner}, \citenamefont {Holec},\ and\
  \citenamefont {Mayrhofer}}]{koutna2021high}%
  \BibitemOpen
  \bibfield  {author} {\bibinfo {author} {\bibfnamefont {N.}~\bibnamefont
  {Koutn{\'a}}}, \bibinfo {author} {\bibfnamefont {A.}~\bibnamefont {Brenner}},
  \bibinfo {author} {\bibfnamefont {D.}~\bibnamefont {Holec}},\ and\ \bibinfo
  {author} {\bibfnamefont {P.~H.}\ \bibnamefont {Mayrhofer}},\ }\bibfield
  {title} {\bibinfo {title} {High-throughput first-principles search for
  ceramic superlattices with improved ductility and fracture resistance},\
  }\href@noop {} {\bibfield  {journal} {\bibinfo  {journal} {Acta Materialia}\
  }\textbf {\bibinfo {volume} {206}},\ \bibinfo {pages} {116615} (\bibinfo
  {year} {2021})}\BibitemShut {NoStop}%
\bibitem [{\citenamefont {Sangiovanni}\ \emph {et~al.}(2010)\citenamefont
  {Sangiovanni}, \citenamefont {Chirita},\ and\ \citenamefont
  {Hultman}}]{sangiovanni2010electronic}%
  \BibitemOpen
  \bibfield  {author} {\bibinfo {author} {\bibfnamefont {D.~G.}\ \bibnamefont
  {Sangiovanni}}, \bibinfo {author} {\bibfnamefont {V.}~\bibnamefont
  {Chirita}},\ and\ \bibinfo {author} {\bibfnamefont {L.}~\bibnamefont
  {Hultman}},\ }\bibfield  {title} {\bibinfo {title} {Electronic mechanism for
  toughness enhancement in {Ti$_x$M$_{1-x}$N} ({M}$=${Mo} and {W})},\
  }\href@noop {} {\bibfield  {journal} {\bibinfo  {journal} {Physical Review
  B}\ }\textbf {\bibinfo {volume} {81}},\ \bibinfo {pages} {104107} (\bibinfo
  {year} {2010})}\BibitemShut {NoStop}%
\bibitem [{\citenamefont {Balasubramanian}\ \emph {et~al.}(2018)\citenamefont
  {Balasubramanian}, \citenamefont {Khare},\ and\ \citenamefont
  {Gall}}]{balasubramanian2018valence}%
  \BibitemOpen
  \bibfield  {author} {\bibinfo {author} {\bibfnamefont {K.}~\bibnamefont
  {Balasubramanian}}, \bibinfo {author} {\bibfnamefont {S.~V.}\ \bibnamefont
  {Khare}},\ and\ \bibinfo {author} {\bibfnamefont {D.}~\bibnamefont {Gall}},\
  }\bibfield  {title} {\bibinfo {title} {Valence electron concentration as an
  indicator for mechanical properties in rocksalt structure nitrides, carbides
  and carbonitrides},\ }\href@noop {} {\bibfield  {journal} {\bibinfo
  {journal} {Acta Materialia}\ }\textbf {\bibinfo {volume} {152}},\ \bibinfo
  {pages} {175} (\bibinfo {year} {2018})}\BibitemShut {NoStop}%
\bibitem [{\citenamefont {Chen}\ \emph {et~al.}(2008)\citenamefont {Chen},
  \citenamefont {Fu}, \citenamefont {Kr{\v{c}}mar},\ and\ \citenamefont
  {Painter}}]{chen2008electronic}%
  \BibitemOpen
  \add{
  \bibfield  {author} {\bibinfo {author} {\bibfnamefont {X.-Q.}\ \bibnamefont
  {Chen}}, \bibinfo {author} {\bibfnamefont {C.~L.}\ \bibnamefont {Fu}},
  \bibinfo {author} {\bibfnamefont {M.}~\bibnamefont {Kr{\v{c}}mar}},\ and\
  \bibinfo {author} {\bibfnamefont {G.~S.}\ \bibnamefont {Painter}},\
  }\bibfield  {title} {\bibinfo {title} {Electronic and structural origin of
  ultraincompressibility of 5$d$ transition-metal diborides {MB$_2$} ({M$=$W,
  Re, Os})},\ }\href@noop {} {\bibfield  {journal} {\bibinfo  {journal}
  {Physical Review Letters}\ }\textbf {\bibinfo {volume} {100}},\ \bibinfo
  {pages} {196403} (\bibinfo {year} {2008})}\BibitemShut {NoStop}%
  }
\bibitem [{\citenamefont {Wang}\ \emph {et~al.}(2019)\citenamefont {Wang},
  \citenamefont {Fu}, \citenamefont {Legut}, \citenamefont {Wei}, \citenamefont
  {Germann},\ and\ \citenamefont {Zhang}}]{wang2019designing}%
  \BibitemOpen
  \add{
  \bibfield  {author} {\bibinfo {author} {\bibfnamefont {N.}~\bibnamefont
  {Wang}}, \bibinfo {author} {\bibfnamefont {Z.}~\bibnamefont {Fu}}, \bibinfo
  {author} {\bibfnamefont {D.}~\bibnamefont {Legut}}, \bibinfo {author}
  {\bibfnamefont {B.}~\bibnamefont {Wei}}, \bibinfo {author} {\bibfnamefont
  {T.~C.}\ \bibnamefont {Germann}},\ and\ \bibinfo {author} {\bibfnamefont
  {R.}~\bibnamefont {Zhang}},\ }\bibfield  {title} {\bibinfo {title} {Designing
  ultrastrong 5$d$ transition metal diborides with excellent stability for
  harsh service environments},\ }\href@noop {} {\bibfield  {journal} {\bibinfo
  {journal} {Physical Chemistry Chemical Physics}\ }\textbf {\bibinfo {volume}
  {21}},\ \bibinfo {pages} {16095} (\bibinfo {year} {2019})}\BibitemShut
  {NoStop}%
  }
\bibitem [{\citenamefont {Burdett}\ \emph {et~al.}(1986)\citenamefont
  {Burdett}, \citenamefont {Canadell},\ and\ \citenamefont
  {Miller}}]{burdett1986electronic}%
  \BibitemOpen
  \add{
  \bibfield  {author} {\bibinfo {author} {\bibfnamefont {J.~K.}\ \bibnamefont
  {Burdett}}, \bibinfo {author} {\bibfnamefont {E.}~\bibnamefont {Canadell}},\
  and\ \bibinfo {author} {\bibfnamefont {G.~J.}\ \bibnamefont {Miller}},\
  }\bibfield  {title} {\bibinfo {title} {Electronic structure of
  transition-metal borides with the {AlB$_2$} structure},\ }\href@noop {}
  {\bibfield  {journal} {\bibinfo  {journal} {Journal of the American Chemical
  Society}\ }\textbf {\bibinfo {volume} {108}},\ \bibinfo {pages} {6561}
  (\bibinfo {year} {1986})}\BibitemShut {NoStop}%
  }
\end{thebibliography}
